%% file: main.tex
\documentclass[sigconf]{acmart}

\AtBeginDocument{%
  \providecommand\BibTeX{{%
    \normalfont B\kern-0.5em{\scshape i\kern-0.25em b}\kern-0.8em\TeX}}}

\acmSubmissionID{6448}

\usepackage{soul}
\usepackage[linesnumbered,ruled,vlined]{algorithm2e}
\usepackage{comment}
\usepackage{multirow}
\usepackage{tabularray}
\newcommand{\RNum}[1]{\uppercase\expandafter{\romannumeral #1\relax}}
\usepackage{xcolor}
\usepackage{tabularx}
\usepackage{subcaption}
\usepackage{array}
\usepackage{colortbl}
\usepackage{booktabs}
\usepackage{graphicx}
\usepackage{array}
\usepackage{booktabs}
\usepackage{balance}

\definecolor{B}{RGB}{33,102,172}
\definecolor{O}{RGB}{179,88,6}
\definecolor{G}{RGB}{127,188,65}

\newcommand{\rr}[1]{#1}

\copyrightyear{2025}
\acmYear{2025}
\setcopyright{cc}
\setcctype{by}
\acmConference[CHI '25]{CHI Conference on Human Factors in Computing Systems}{April 26-May 1, 2025}{Yokohama, Japan}
\acmBooktitle{CHI Conference on Human Factors in Computing Systems (CHI '25), April 26-May 1, 2025, Yokohama, Japan}\acmDOI{10.1145/3706598.3714280}
\acmISBN{979-8-4007-1394-1/25/04}

\definecolor{purple}{HTML}{7d41a8}
\definecolor{yellow}{HTML}{fbcb28}
\definecolor{fuchsia}{HTML}{ff3ac9}
\definecolor{blue}{HTML}{0000FF}
\definecolor{brown}{HTML}{ab762b}
\definecolor{red}{HTML}{ff0000}

\newcommand{\toolname}[0]{\textsc{Patrika}}

\newcommand{\pheading}[1]{\vspace{4px}\noindent\textbf{#1}}

\newenvironment{tight_itemize}{\begin{itemize} \itemsep -1pt}{\end{itemize}}

\begin{document}

\title{AI-Enabled Conversational Journaling for Advancing Parkinson's Disease Symptom Tracking}

\author{Mashrur Rashik}
\affiliation{
  \institution{University of Massachusetts Amherst}
  \city{Amherst}
  \state{Massachusetts}
  \country{USA}
}
\email{mrashik@cs.umass.edu}

\author{Shilpa Sweth}
\affiliation{
  \institution{University of Massachusetts Amherst}
  \city{Amherst}
  \state{Massachusetts}
  \country{USA}
}
\email{ssweth@cs.umass.edu}

\author{Nishtha Agrawal}
\affiliation{
  \institution{New York University}
  \city{New York City}
  \state{New York}
  \country{USA}
}
\email{na3533@nyu.edu}

\author{Saiyyam Kochar}
\affiliation{
  \institution{University of Massachusetts Amherst}
  \city{Amherst}
  \state{Massachusetts}
  \country{USA}
}
\email{skochar@umass.edu}

\author{Kara M Smith}
\affiliation{
  \institution{University of Massachusetts Chan Medical School}
  \city{Worcester}
  \state{Massachusetts}
  \country{USA}
}
\email{kara.smith@umassmemorial.org}

\author{Fateme Rajabiyazdi}
\affiliation{
  \institution{Carleton University}
  \city{Ottawa}
  \state{Ontario}
  \country{Canada}
}
\email{fateme.rajabiyazdi@carleton.ca}

\author{Vidya Setlur}
\affiliation{
  \institution{Tableau Research}
  \city{Palo Alto}
  \state{California}
  \country{USA}
}
\email{vsetlur@tableau.com}

\author{Narges Mahyar}
\affiliation{
  \institution{University of Massachusetts Amherst}
  \city{Amherst}
  \state{Massachusetts}
  \country{USA}
}
\email{nmahyar@cs.umass.edu}

\author{Ali Sarvghad}
\affiliation{
  \institution{University of Massachusetts Amherst}
  \city{Amherst}
  \state{Massachusetts}
  \country{USA}
}
\email{asarv@cs.umass.edu}

\renewcommand{\shortauthors}{Mashrur Rashik et al.}

\begin{abstract}
\input{new_sections/0-abstract}

\end{abstract}

\ccsdesc[500]{Human-centered computing~Interactive systems and tools}
\ccsdesc[500]{Human-centered computing~Natural language interfaces}
\keywords{Conversational implicature, Gricean maxims, context, journaling, healthcare.}

\begin{teaserfigure}
  \centering
  \includegraphics[width=0.8\textwidth]{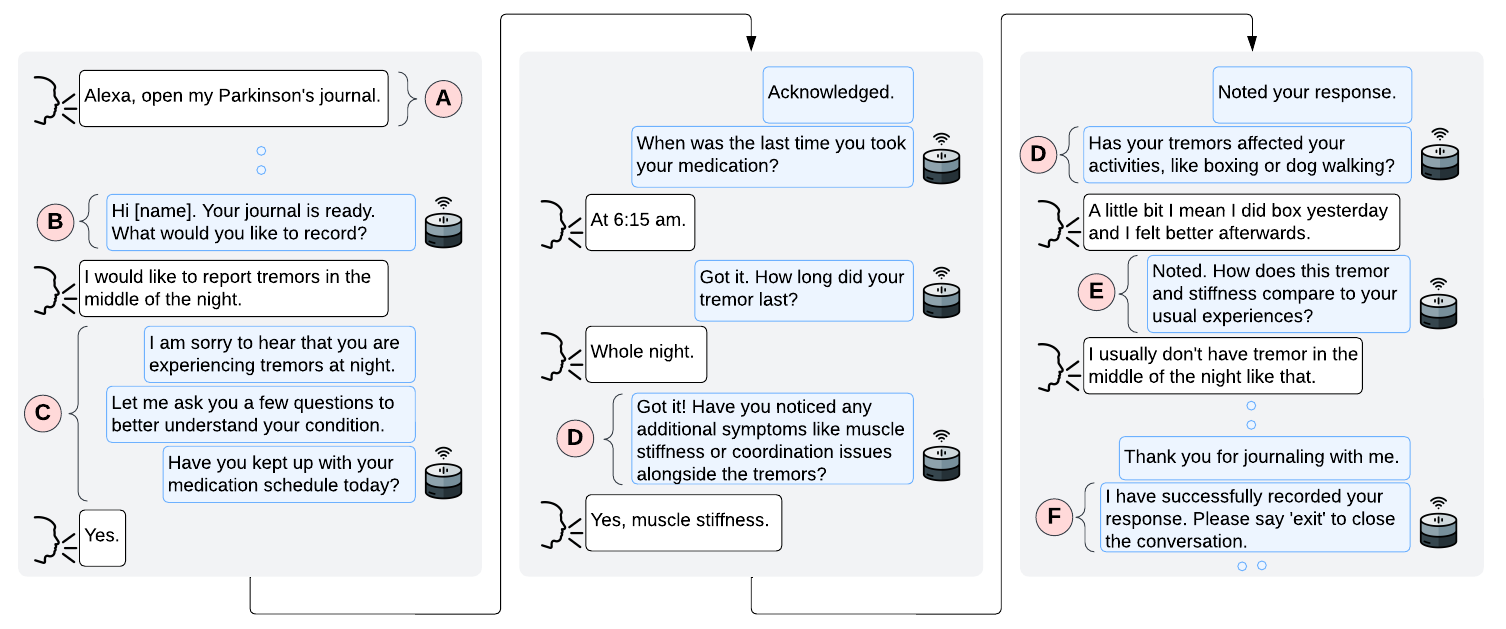}
  \caption{This figure shows an anonymized snippet of conversation between a patient with Parkinson's disease and \toolname{}. Here:
  (A) The patient invokes their journal using a voice command to initiate the conversation. 
  (B) \toolname{} responds with a personalized greeting [name] and prompts the patient to start recording.
  (C) \toolname{} detects ``tremor'' in the patient's previous entry and initiates collecting additional data via asking relevant follow-up questions.
  (D) \toolname{} utilizes the conversation history with the patient to personalize the follow-up question, asking about symptoms previously co-occurring with tremors and their impact on their routine activities.
  (E) \toolname{} utilizes conversation history to identify and compare the severity of recurring symptoms. For instance, in a previous entry, the patient reported experiencing tremors with no change in severity or overall experience up to that point. \toolname{} compares this information to the most recent tremor episode, which occurred in the middle of the night.
  (F) \toolname{} confirms that the data has been recorded and then prompts the user to either end the conversation or initiate a new one.
  }
  \label{fig:teaser}
\end{teaserfigure}

\maketitle

\input{new_sections/1-introduction}

\input{new_sections/2-related_work}

\input{new_sections/3-system-design}

\input{new_sections/4-experiment_1}

\input{new_sections/5-system-improvements}

\input{new_sections/6-experiment_2}

\input{new_sections/7-discussion}

\input{new_sections/8-limitations}

\balance

\input{new_sections/9-conclusion}

\bibliographystyle{ACM-Reference-Format}
\bibliography{bibliography}

\end{document}

%% file: new_sections/0-abstract.tex
Journaling plays a crucial role in managing chronic conditions by allowing patients to document symptoms and medication intake, providing essential data for long-term care. While valuable, traditional journaling methods often rely on static, self-directed entries, lacking interactive feedback and real-time guidance. This gap can result in incomplete or imprecise information, limiting its usefulness for effective treatment. To address this gap, we introduce \toolname{}, an AI-enabled prototype designed specifically for people with Parkinson's disease (PwPD).
The system incorporates cooperative conversation principles, clinical interview simulations, and personalization to create a more effective and user-friendly journaling experience. Through two user studies with PwPD and iterative refinement of \toolname{}, we demonstrate conversational journaling's significant potential in patient engagement and collecting clinically valuable information. Our results showed that generating probing questions \toolname{} turned journaling into a bi-directional interaction. Additionally, we offer insights for designing journaling systems for healthcare and future directions for promoting sustained journaling.

%% file: new_sections/1-introduction.tex
\section{Introduction}

Conversational journaling is a technique for collecting patient-generated health data, where patients interact with a digital system using natural language to document their health and daily experiences, such as symptoms, medication intake, and physical and mental states. Journaling practice has shown promise in supporting long-term care, especially for individuals with chronic medical conditions~\cite{shapiro2012patient, sohal2022efficacy}, as this approach enhances patient-healthcare provider communication, supports more accurate diagnosis, improves treatment planning, and ultimately leads to better health outcomes~\cite{baikie2005emotional, ashburn2008circumstances, winslow2018self, milne2022self, Mishra2019}. Despite the longstanding presence of conversational systems for health data collection (e.g.,~\cite{shortliffe1975model, miller1985internist}), early systems were limited by their ability to understand and generate natural language, which constrained their effectiveness in real-world applications. However, recent advancements in natural language processing (NLP) have opened new possibilities for developing sophisticated conversational journaling systems~\cite{achiam2023gpt, salemi2024optimization}. 

Prior work indicates the feasibility of conversational journaling as a journaling method~\cite{maharjan2021can}, factors influencing sustained adaptation by patients~\cite{Maharjan2022ExperiencesOA}, and its utility for enabling self-reflection in mental health context~\cite{kim2023mindfuldiary}.
These systems can potentially recreate essential elements of the ``clinical interview,'' a process that healthcare professionals utilize to gather in-depth information about patients' physical and mental health through targeted Q\&A~\cite{kim2023mindfuldiary}. Moreover, coupling conversational journaling with voice interaction can reduce accessibility barriers for individuals with motor and visual impairments ~\cite{maharjan2021can, Maharjan2022ExperiencesOA}.

Despite this potential, there is still a notable lack of empirical understanding of designing effective conversational journaling systems for better interaction with individuals having chronic conditions. To address this gap, our interdisciplinary research team — including a neurologist specializing in movement disorders and patients with Parkinson's disease, an expert in engineering and implementation of conversational agents, a specialist in the linguistic design of conversational AI agents, two human-computer interaction experts, and a health informatics specialist— undertook an exploratory investigation in which we designed and developed \toolname{}~\footnote{\toolname{} originated from Sanskrit that translates to ``journal'' or ``magazine.''}, an AI-enabled conversational journaling prototype for People with Parkinson's disease (PwPD). PwPD can significantly benefit from regularly journaling their symptoms and medication~\cite{vafeiadou2021self, nunes2019agency, mcnaney2020future}. Parkinson's disease symptoms frequently vary and fluctuate throughout the day, necessitating frequent adjustments to typically complex medication regimens. While medications may alleviate specific symptoms, side effects can also trigger or exacerbate others. For instance, long-term use of Levodopa, one of the most effective medications for managing motor symptoms in Parkinson's disease, can lead to the development of dyskinesia, characterized by involuntary and erratic body movements~\cite{thanvi2007levodopa}. Therefore, collecting and communicating detailed information about symptom occurrences, fluctuations, and medication effects is essential for tailoring treatment plans and optimizing symptom management in Parkinson's disease~\cite{mcnaney2020future, vafeiadou2021self, nunes2018understanding}.

Research shows that people with Parkinson's disease frequently seek personalized and descriptive methods to self-track their daily lived experiences~\cite{vafeiadou2021self}. While sensing technologies embedded in tools such as wearables and smartphones for symptom tracking have expanded rapidly in recent years, much of this technology focuses on quantitative rather than qualitative data collection and is primarily centered on motor symptoms rather than non-motor symptoms. As a result, these tools may fail to align with patients' perspectives, which prioritize factors that significantly impact their daily experiences and quality of life~\cite{mcnaney2020future, skorvanek2018relationship}. Tools such as diary apps and patient portals support qualitative data collection by enabling patients to express and record their experiences. However, these tools are predominantly designed to passively receive and record information from the patient, missing the opportunity to transform journaling into a bi-directional exchange between the patient and the system with the objective of enriching data collection. Additionally, PwPD face unique challenges that often render current journaling tools less effective. Motor symptoms such as tremors, rigidity, and bradykinesia (slowness of movement) can make writing or typing difficult, while non-motor symptoms—including cognitive decline, speech difficulties, and emotional changes further complicate the accurate documentation of daily health experiences. This combination of factors highlights the need for personalized technologies that support individual patient agency and autonomy~\cite{nunes2019agency}. Voice-enabled systems, particularly those enhanced by AI-driven conversational capabilities, offer a promising solution by reducing the physical strain of data entry and improving the accuracy of symptom tracking through dynamic, interactive dialogues. 

We designed \toolname{} as a personalized conversational journaling prototype that supports the self-tracking of a wide variety of motor and non-motor experiences of daily living in PwPD. The prototype was developed based on clinical management practices that PwPD would typically undertake with their clinician, to help bring key components of the clinical interview process into a home-based and individualized form. Upon detecting Parkinson's disease-related symptoms (e.g., tremor) in patients' utterances, \toolname{} generates follow-up questions that probe relevant aspects such as the duration of the symptom and medication intake (see Figure~\ref{fig:teaser}). The specific symptoms and corresponding topics for these questions were selected based on recommendations from our team’s neurologist (\autoref{tab:sympthoms_and_topics}). 
Through iterative design and development of \toolname{}, along with two studies involving PwPD, our main contribution lies in validating the potential and benefits of AI-enabled interactive journaling. Specifically, this work makes two primary contributions:


\begin{itemize}
    \item \textbf{Artifact contribution:} Designed and developed \toolname{}, a novel prototype system that advances state-of-the-art by integrating cooperative conversation principles, dialogue personalization and empathy into conversation design, fostering natural and engaging interactions. The system also adapts essential elements from clinical interviews to improve journaling aimed at enriching data collection from PwPD.
    \item \textbf{Empirical \rr{contribution}:} Demonstrated that conversational journaling supported by Large Language Models (LLMs) can effectively identify user intents and generate relevant, personalized follow-up questions. Through iterative improvements, \toolname{} achieved 99\% intent identification accuracy and an 81\% success rate in generating personalized responses. \rr{Additionally, our qualitative analysis of journaling sessions and post-study interviews revealed that PwPD found} \toolname{} \rr{intuitive and engaging, with follow-ups that facilitated the disclosure of clinically rich insights about PwPD symptoms, medication routines, and daily experiences. Feedback from two movement disorder specialists further highlighted} \toolname{}’s \rr{ability to capture nuanced information with the potential to direct patient care.}
\end{itemize}

%% file: new_sections/2-related_work.tex
\section{Related Work}
Our work builds upon three themes of research: natural language interfaces, conversational agents for journaling, and journaling in healthcare. 

\subsection{Natural language interfaces}
Prior work has emphasized the importance of maintaining a natural conversational flow 
for effective human-agent interaction~\cite{skantze2021turn, jiang2023communitybots}. One of the challenges for building a natural conversational flow is the accurate identification of user intention that can lead to a more natural and smoother dialogue between a user and a chatbot agent~\cite{zhang2018making}. The accurate identification of user intention during a conversation includes both the user's intent to respond as well as their refusal to respond by demonstrating unwillingness towards the current conversation topic~\cite{xiao2020if, adamopoulou2020overview}.
Lack of effective user intent identification can result in user disengagement and dissatisfaction with the conversational agent~\cite{see2019makes, xiao2020if, han2021designing}. Another challenge is to identify the explicit and implicit cues from the user to understand when to follow up with a response.
In real-world conversations, people often rely on explicit cues, such as verbal requests, and implicit cues, such as non-verbal body and facial expressions, to take turns between speakers and transition between topics~\cite{wiemann2017turn}. Although natural language understanding (NLU) could identify explicit and implicit cues, simulating a natural conversation between a chatbot and a human remains challenging due to other nuances, such as identifying the presence of metaphors, idioms, sarcasm, or rhetorical questions ~\cite{jhamtani2021investigating, nordberg2019designing}.
Furthermore, machine learning and natural language research have explored automated machine learning methods to understand and adapt to a user's conversational style, such as the usage of internet phonographs and jargon to develop user-specific data for training the conversational agent~\cite{lee2020human}. 
However, building conversational agents capable of adapting to user patterns is computationally expensive~\cite{cuayahuitl2019ensemble} and often requires validation from users and effective human-in-the-loop processes to avoid mistrust, confusion, and dissatisfaction with the conversational agent~\cite{fernau2022towards}. 

\toolname{} focuses on the unique challenges of user intent identification and conversational flow specifically for PwPD.  The system is designed to accurately interpret the often complex and nuanced symptoms reported by PwPD, such as variations in tremor intensity, medication effects, and daily fluctuations in motor function. Additionally, the system incorporates a personalized and adaptive conversational style by employing Retrieval-augmented Generation (RAG), an NLP technique to enhance the generation of responses by dynamically tailoring follow-up questions based on the user's conversational history and specific health context.

\subsection{Conversational Agents for Journaling}

Conversational systems have become increasingly common across various domains, including e-commerce, mental health, and education~\cite{folstad:2017,Provoost2017EmbodiedCA,Winkler2018UnleashingTP, Maharjan2022ExperiencesOA}. Traditionally a solitary activity, journaling can be enhanced by conversational systems, offering users additional support in areas such as self-reflection, engagement, and problem-solving~\cite{kocielnik2018designing}.

For example, Lucas et al.~\cite{lucas:2014} found that people are more willing to disclose personal information to a conversational system, highlighting the potential for deeper self-reflection in journaling. Conversational systems also provide adaptive prompting mechanisms to help with the journaling process. 
For instance, Miner et al.~\cite{miner:2016} observed that intelligent prompting can encourage users to explore unexamined emotional areas or topics, thereby enhancing the quality of self-reflection and the journaling experience as a whole.
Seo et al.~\cite{seo2024chacha} created a conversational agent that prompts children to share their emotions. 
An exploratory study using this system demonstrated that conversational systems can effectively recognize children's intentions and associated emotions, allowing for more empathetic conversations.

Our work addresses the unique challenges faced by PwPD in the context of health journaling. While previous conversational systems have been successful in enhancing engagement and self-reflection, they often lack the nuanced understanding required to effectively support users with complex chronic conditions. \toolname{} leverages LLMs for intent recognition and RAG for personalized prompts to tailor the journaling experience specifically to the needs of PwPD, adapting its responses based on the user's ongoing health status and historical data.

\subsection{Journaling in Healthcare}
The interest in collecting patient-generated health data has been growing due to its benefits for enhancing patient self-awareness and health monitoring as well as providing valuable context for diagnosis and clinical decision-making~\cite{Figueiredo2020PGD, Lordon2020, Mishra2019, Chung2019, Rajabiyazdi2021}. The act of self-expression through journaling can also offer psychological benefits, such as stress reduction~\cite{ullrich:2002, smyth:1999}. In the specific context of Parkinson's, research highlights several benefits of journaling, including PwPD's improved ability to self-reflect and their increased engagement in the clinical decision-making process~\cite{nunes2019agency, nunes2018understanding, vafeiadou2021self}. A variety of technologies enable data collection for PwPD. Sensing tools like wearable devices gather objective quantitative data such as metrics on gait and the amplitude and frequency of tremors~\cite{mcnaney2020future, adams2023using, zhang2019pdmove}. Diary apps and patient portals allow PwPD to express and record their subjective experiences in natural language~\cite{maharjan2021can, rahman2024user}. Yet, these tools predominantly use a passive journaling model, only receiving and recording patients' input. Conversational journaling is an alternative model in which an AI-enabled agent engages the patient in dialogue, transforming journaling into a dynamic bi-directional interaction~\cite{jovanovic2020chatbots, Lee2020, Sheth2019}. This opens up opportunities for expanding the depth and breadth of data collection via targeted, personalized conversation with PwPD~\cite{nunes2019agency, mcnaney2020future}. 

Prior research highlights the benefits of conversational journaling across various contexts, including enhancing self-disclosure among mental health patients with their specialists~\cite{Lee2020}, supporting help-seeking behaviors in college students dealing with anxiety and depression~\cite{Fitzpatrick2017}, and facilitating self-monitoring for cancer patients~\cite{Lu2021}. Conversational agents have also been widely applied in healthcare for healthy living~\cite{kocaballi2020responses}, medical information seeking~\cite{yang2021clinical, miner:2016}, and chronic condition management~\cite{vaidyam:2019, zand2020exploration, hong2021voice, vaidyam:2019}. Inkster et al.~\cite{inkster:2018} showed that empathy-driven conversational agents could enhance mental well-being interventions. Recent advances in LLMs for conversational agents have focused on generating relevant, consistent, and timely follow-up interactions~\cite{li2024beyond, xygkou2024mindtalker, nepal2024contextual}, and have highlighted the potential for personalized interactions in healthcare~\cite{xygkou2024mindtalker, jo2024understanding}.

Developing conversational journaling for PwPD is a complex task. It requires addressing numerous technical and human-centered design challenges, such as accurately and promptly identifying issues, generating relevant follow-up questions tailored to the patient’s profile and history, and considering linguistic nuances that impact patient experience and their perception of the system's value and utility. To the best of our knowledge, our work is among the first to empirically investigate these challenges through the design, development, and real-world evaluation of \toolname{} with Parkinson’s patients.

%% file: new_sections/3-system-design.tex
\section{\toolname{} Design and Implementation}\label{sec:system_design}
Developing a journaling system capable of simulating and sustaining targeted, relevant, and relatable conversations with a patient about their health is a complex and multidisciplinary endeavor involving aspects of natural language conversational paradigms~\cite{valizadeh2022ai}, healthcare~\cite{nissen2022effects, maharjan2021can}, and human-computer interaction~\cite{setlur2022you}. To help guide the development and evaluation of \toolname{}, we focused on addressing the following key research questions:

\begin{itemize}
\item \textbf{RQ1}: How can we effectively simulate the clinical interview process through targeted, relevant, and personalized questions?
\item \textbf{RQ2}: How can we solicit responses from PwPD that facilitate rich data collection and offer clinical insights, deepening the understanding of patient experiences and disease management?
\item \textbf{RQ3}:  How do PwPD evaluate the usefulness and effectiveness of conversational journaling for tracking and managing their condition?
\end{itemize}

\begin{table*}[t]
    \centering
    \caption{This table lists selected Parkinson's symptoms with corresponding follow-up topics and example questions. Colored boxes show connections between probing topics and symptoms. When \toolname{} detects a symptom, it uses these topics to generate follow-up questions. The mapping between symptoms and probing topics is not one-to-one; for instance, a question about ``duration'' would be asked for ``tremor'' but not for ``falling.''}
    \includegraphics[width=\linewidth]{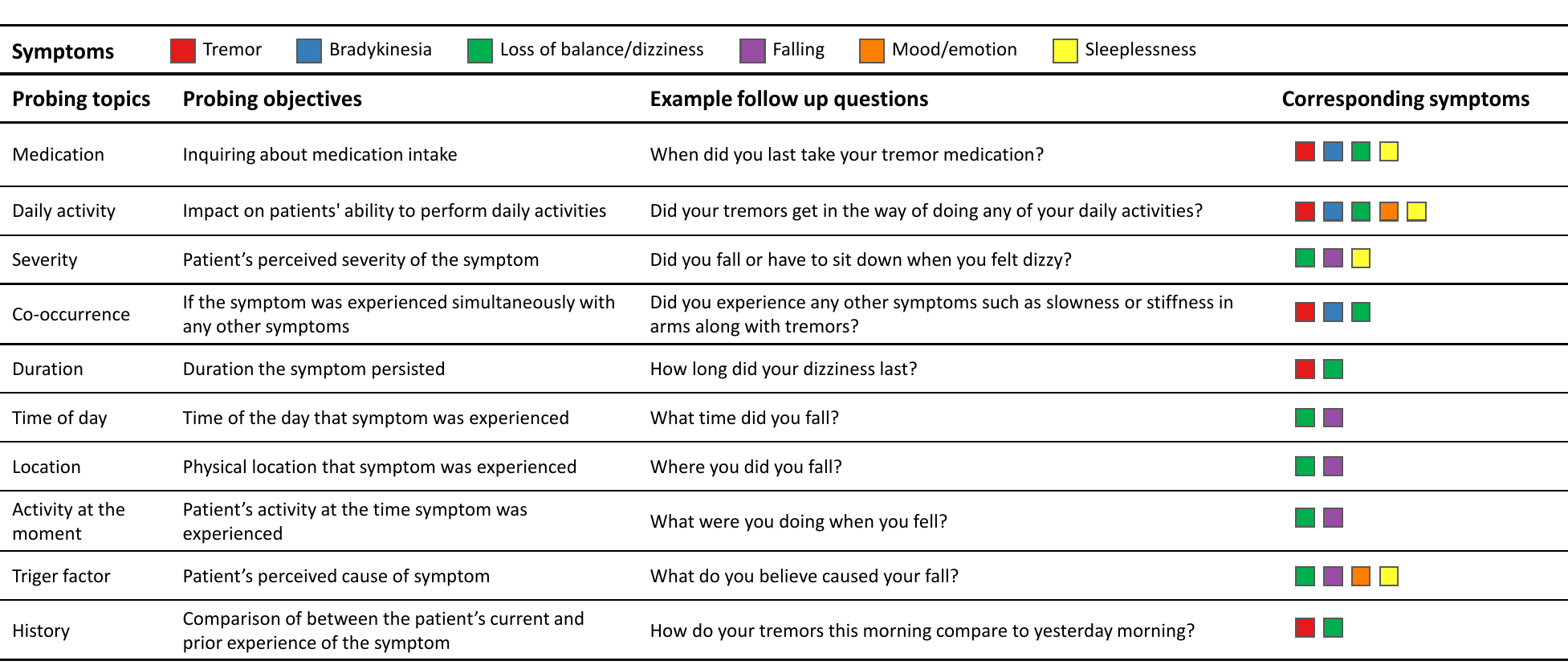}
    \label{tab:sympthoms_and_topics}
\end{table*}

\subsection{Core Design Considerations and Criteria}
Drawing on prior research~\cite{setlur2022you, jovanovic2020chatbots} and extensive team discussions, we identified three core considerations for designing \toolname{}: 1) adhering to cooperative conversation principles between AI and users, 2) simulating a clinical discussion through timely, relevant, and personalized follow-up questions, and 3) enhancing accessibility and ease of use. The following sections detail how these principles were transformed into actionable design considerations.

\subsubsection{Adherence to tenets of cooperative conversation between AI agents and people}
Gricean Maxims~\cite{grice1975} offers a framework for guiding the design of an AI agent's language and conversational behavior~\cite{setlur2022you}. The four maxims—\textit{Quantity}, \textit{Quality}, \textit{Relevance}, and \textit{Manner}—describe how speakers should act cooperatively to ensure mutual understanding in effective communication (see supplementary materials). Building on prior work~\cite{setlur2022you}, we developed design criteria (DC) 1 to 4 to govern the journaling agent's conversational behavior according to these maxims. These criteria were informed by the need to ensure that interactions adhere to Gricean conversational principles, such as relevance and clarity, while also addressing the specific challenges faced by PwPD, including motor and cognitive impairments that affect their ability to engage in traditional journaling methods.

\begin{tight_itemize}
    \item \noindent\textbf{DC-1: Provide clear conversation initiation \& termination.}  The conversational agent's greeting and termination should clearly signal when to start and stop recording symptoms (Maxims of Quality and Manner).
 
    \item \noindent\textbf{DC-2: Provide relevant acknowledgements.} The agent should give relevant and accurate acknowledgments (Maxims of Quality, Relevance, and Manner).

    \item \noindent\textbf{DC-3: Support effective transitions between topics.} The agent should manage topic-switching smoothly, confirming with the user before transitioning (Maxims of Relevance and Manner)~\cite{skantze2021turn, jiang2023communitybots}.

     \item \noindent\textbf{DC-4: Provide conversation repair \& refinement.} The system should promptly address conversation issues and improve interaction quality by clarifying ambiguities and resolving misunderstandings~\cite{setlur2022you, chan2022challenges}.

\end{tight_itemize}
Design criteria DC1-4 apply to most conversational agents, but health-related conversations require additional considerations, leading to four additional design criteria:

\begin{tight_itemize}
    \item \noindent\textbf{DC-5: Implement security measures.} The conversational agent must implement strong privacy measures, including end-to-end encryption, to protect sensitive user entries~\cite{li2023security}.

    \item \noindent\textbf{DC-6: Integrate empathy in the agent responses.} 
    The conversational agent should incorporate empathy in their responses (Maxim of Manner)~\cite{adikari2022empathic,daher2020empathic,devaram2020empathic}. 
       
    \item \noindent\textbf{DC-7: Understand medical terminology.} The agent should recognize medical and lay terms related to the health domain, in our case Parkinson's disease~\cite{safi2020technical}.

 \item \noindent\textbf{DC-8: Refrain from providing healthcare advice.} The conversational agent should refrain from providing explicit or implicit clinical or general healthcare advice to the patient~\cite{daher2020empathic, palanica2019physicians}.
    
\end{tight_itemize}

\subsubsection{Effectively Simulating the clinical interview process}

Personalizing conversations by using the patient's name and referencing their medical history builds trust and increases relevance~\cite{liu2022roles, liu2022effects, kocaballi2019personalization}.
 
  \begin{tight_itemize}

         \item \noindent\textbf{DC-9: Generate relevant and timely follow-up questions.} The questions should be contextually relevant and timely, aiming to elicit additional patient information.

         \item \noindent\textbf{DC-10: Support personalization.} 
              The agent should use dialogue history to personalize the content and the delivery of the conversation.
    
    \end{tight_itemize}

The inclusion of \textbf{DC-9} led to two key questions: (1) What expressions in PwPD utterances should trigger follow-up questions? and (2) What additional information would enhance the clinical value of journaling? To answer these questions, we collaborated with our team's neurologist, identifying six common Parkinson's symptoms—\textit{tremors}, \textit{bradykinesia}, \textit{dizziness}, \textit{falls}, \textit{mood/emotion}, and \textit{sleeplessness}—as triggers for follow-ups. Focusing on these symptoms ensures relevance for a broad range of PwPD. The neurologist also recommended ten probing topics, including \textit{medication}, \textit{daily activity}, and \textit{duration}, to gather more nuanced data. For symptoms outside the pre-selected set or other patient issues, \toolname{} records and acknowledges them without issuing follow-ups. Table~\ref{tab:sympthoms_and_topics} details the symptoms, probing topics, example questions, and their interrelationships.

\subsubsection{Enhanced accessibility and ease of use}
The final design criteria focus on making \toolname{} accessible, particularly for PwPD with motor function issues. This approach aims to accommodate users who may struggle with graphical user interface (GUI)-based or physical data entry, ensuring \toolname{} remains user-friendly and accessible for use.

\begin{itemize}
 \item \noindent\textbf{DC-11: Enable low-barrier and accessible journaling support.} Journaling should be accessible and designed to require minimal physical effort for PwPD.

\end{itemize}

Examples of how these design considerations can be operationalized in agent-human conversations are presented in the supplementary materials.

\subsection{System Implementation}

\begin{figure*}[t]
    \centering
    \includegraphics[width=\linewidth]{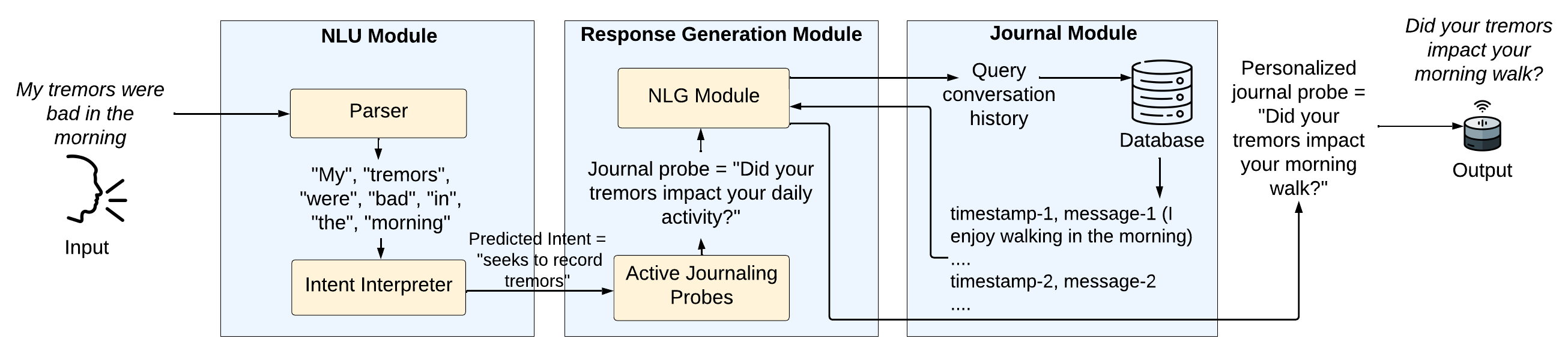}
    \caption{This figure shows the system overview of \toolname{}. The system comprises three modules: Natural Language Understanding (NLU), response generation, and journal. The NLU module tokenizes the user input and predicts the user's intent. This predicted intent is then passed on to the response generation module, which uses both the intent and the conversation context to predict a relevant journal prompt or response. To personalize the journal prompt, the response generation module queries the journal module, which returns the user's conversation history. The response generation module uses the conversation history to generate a personalized journal prompt, which is then delivered to the user through Alexa.}
    \label{fig:system_overview}
\end{figure*}

We implemented \toolname{} as a voice-enabled system (\textbf{DC-11}) that runs as an API endpoint on Amazon AWS~\cite{awsec2}, accessible via Amazon Alexa devices. The system uses Alexa's speech-to-text and text-to-speech components to process voice input and generate voice responses. We developed \toolname{}'s natural language components using Rasa~\cite{bocklisch2017rasa}, an open-source framework for conversational agents. To ensure data security and privacy, we store NL components and journaling data in Amazon AWS's HIPAA-compliant DynamoDB environment. \toolname{} uses ngrok~\footnote{https://ngrok.com/} to establish a secure HTTPS endpoint for encrypted TLS communication between Alexa and \toolname{} (\textbf{DC-5}).
The TLS encryption ensures that all data sent between Alexa and \toolname{} is encrypted with a key unique to each patient. This protects the patient's data from unauthorized access and ensures confidentiality. Figure~\ref{fig:system_overview} outlines \toolname{}'s three main modules: the \textit{Natural Language Understanding (NLU) Module}, the \textit{Response Generation Module}, and the \textit{Journal Module}.

\subsubsection{Natural Language Understanding (NLU) Module}
\begin{figure*}[t]
    \centering
    \includegraphics[width=\linewidth]{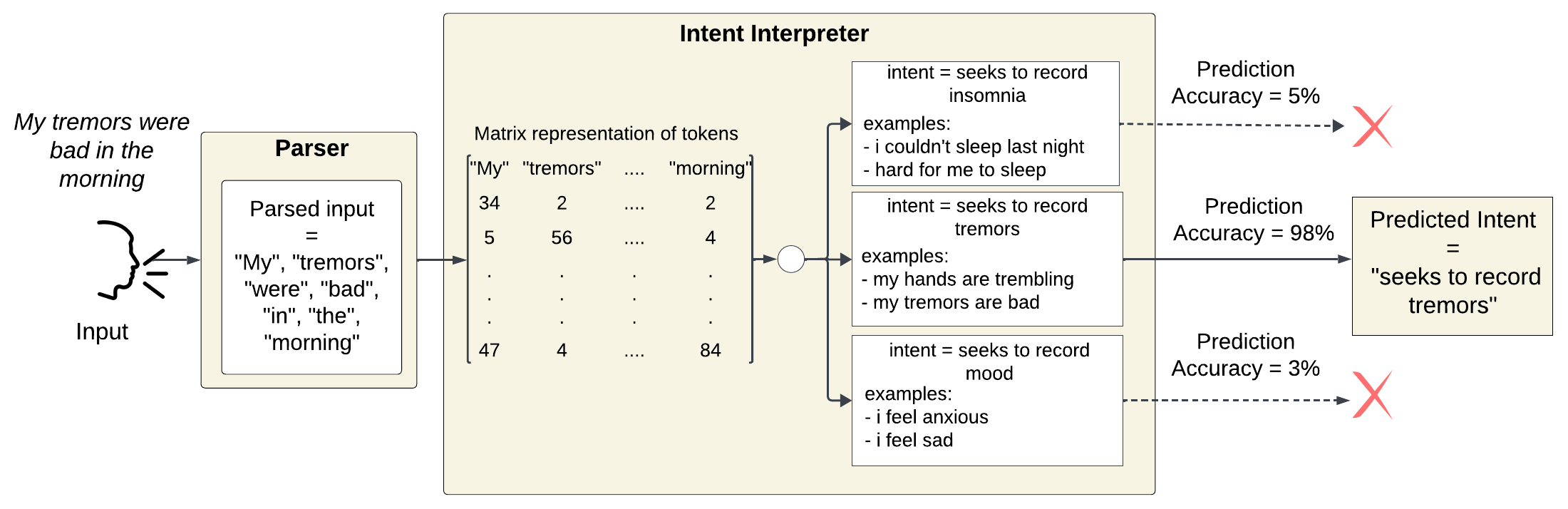}
    \caption{This figure shows the workflow of the Natural Language Understanding (NLU) Module of \toolname{}.
    The NLU module consists of two components: a parser and an intent interpreter. The parser converts the user's input into a string of words, which are then passed to the intent interpreter. The intent interpreter transforms this word sequence into a matrix representation. This matrix is fed into a transformer-based model called the ``DIETClassifier,'' which compares it to a set of predefined intents. The ``DIETClassifier'' then predicts the intent that best matches the user's input.
    }
    \label{fig:nlu_module}
\end{figure*}

The NLU Module processes the user's voice input and converts it to text via Alexa's speech-to-text component before passing the user's intent information to the Response Generation Module. 
This module consists of two main components: \textit{Parser} and an \textit{Intent Interpreter}. The Parser component tokenizes the input into segments of the text typically represented by words, subwords, or punctuations. We use a Spacy parser~\cite{spacynlu} for its speed, accuracy, and ease of use. The parser applies context-free grammar and regular expressions to segment text accurately~\cite{webster1992tokenization}. As shown in Figure~\ref{fig:nlu_module}, the input ``My tremors were bad in the morning'' is tokenized into individual words: \{``My,'' ``tremors,'' ``were,'' ``bad,'' ``in,'' ``the,'' ``morning''\}. This token sequence is then passed to the intent interpreter.

The Intent Interpreter component in the NLU Module maps the parsed tokens to an intent that reflects the user's goal or purpose of communication with \toolname{}. In the system, intents are divided into two main categories (a) Symptom Intents: These intents capture inputs corresponding to the specific Parkinson's symptoms that \toolname{} supports users in journaling, as shown in Table~\ref{tab:sympthoms_and_topics}.
Examples of such intents include the user's desire to journal their tremor (e.g., ``I want to journal my shaky hands''), report their medication intake (e.g., ``\textit{I took my medicine at 10 in the morning}''), or mention a daily activity affected by Parkinson's (e.g., ``\textit{My tremor makes it hard to cook}'').
(b) Conversational Intents: These intents capture inputs that are unrelated to Parkinson's but are essential for normal dialogue management and require additional follow-ups to refine the conversation (DC-4). For example, user inputs like expressing confusion (e.g., ``\textit{I don't understand}''), requests for additional follow-ups (e.g., ``\textit{Can you provide more information?}''), or expressing disinterest (e.g., ``\textit{I don't want to talk about this}''). To ensure that the intent categories are comprehensive and accurate, we reference Rasa's crowd-sourced NLU data~\cite{rasanlu2021git} as a guideline. 

To predict the intent, the token sequence is transformed into a matrix using a pre-trained Spacy model~\cite{spacynlu} that captures the contextual meaning in the user input~\cite{kenton2019bert}. 
The Intent Interpreter uses this matrix to classify an intent category by identifying patterns in the token sequence that match known intents.
To do so, the Intent Interpreter uses a trained classifier model.
In this work, we use the ``DIETClassifier'' from Rasa as the Intent Interpreter due to its robustness and accuracy~\cite{bunk2020diet}.
This classifier is trained on hand-crafted NLU data, with example sentences for each intent category (see supplementary materials). These example sentences represent potential relevant user inputs under each intent category.
We also utilized the Unified Parkinson's Disease Rating Scale (UPDRS)~\cite{movement2003unified} questionnaire, a comprehensive questionnaire used to assess the severity of PwPD, to include medical terms and alternative lay language relevant to PD (\textbf{DC-7}). We trained the intent interpreter for 100 epochs, using 80\% of data as the training set. An epoch is one full run through all the training examples.
The remaining 20\% served as the test set, on which our interpreter achieved a 98\% accuracy.


\subsubsection{Response Generation Module}
The Response Generation Module uses the predicted intent to generate a response for the user, as well as relevant follow-up journal prompts based on the conversation context (\textbf{DC-1}, \textbf{DC-2}, \textbf{DC-3}, \textbf{DC-9}).
For each intent, we create a list of follow-up journal prompts that correspond to the probing topics (see Table~\ref{tab:sympthoms_and_topics}).
\toolname{} replicates the clinical interview process by asking symptom-specific follow-up journal prompts based on the conversation context, similar to how a clinician would delve deeper into a patient's condition to gain valuable insights and refine treatment plans.
To select a journal prompt from this list, the Response Generation Module uses a trained classifier model.
We train this classifier model on our handcrafted dataset, which consists of $1087$ sample journaling interactions between \toolname{} and a patient (see supplementary materials). 
The dataset was iteratively updated by three research team members based on internal testing.
To test the dataset, the team members used 10\% of the dataset as the baseline.
Then, they compare the output from \toolname{} trained on the remaining 90\% of the dataset to the baseline output.
The lead author met with the three members each week and manually marked the responses generated from \toolname{} that did not correspond to the baseline.
The team computed the average number of mismatches and updated the dataset to reduce this value.
The Response Generation Module feeds the conversation context into the trained classifier, providing an accuracy score over every journal prompt relevant to intent. The journal prompt with the highest accuracy is then selected. 
If the highest accuracy is less than 20\%, then the Response Generation Module initiates a conversation repair and refinement process (\textbf{DC-4}).
During this process, the Response Generation Module empathetically (\textbf{DC-6}) acknowledges the reception of the user's input without providing any medical advice (\textbf{DC-8}) and repeats itself. In practice, we found that a threshold of 20\% tended to yield reasonable results.

We use TEDPolicy~\cite{vlasov2019dialogue} from Rasa for its computational efficiency in predicting journal prompts based on user intent. Following standard dataset-splitting practices~\cite{guyon1997scaling}, we trained TEDPolicy on 80\% of our handcrafted dataset for 100 epochs, with the remaining 20\% as the test set. The TEDPolicy model achieved a prediction accuracy of 97.5\% on the test set.

\subsubsection{Journal Module}
We use the Journal Module to read and write conversations between the user and \toolname{}. This module consists of a semi-structured database where the data fields and their types are defined, but the data isn't stored using any fixed structure. 
Each journal entry is cataloged as a timestamped pair with the corresponding user input. The Response Generation Module accesses the conversation history by querying the database using the user’s unique identification number and current journal symptom. This module returns the history that is relevant to the user's current journal symptom. We chose MongoDB~\cite{mongodb} as the database for its flexibility in handling semi-structured data and fast query processing.

\subsubsection{Personalization Mechanism}

After selecting a relevant journal prompt, the Response Generation Module personalizes the prompt using the journal history and the conversation context (\textbf{DC-10}).
The Response Generation Module queries the Journal Module for the user's history, which returns records from previous sessions. These records, combined with the conversation context, are used by the Natural Language Generation (NLG) component to personalize the journal prompt.
NLG uses a prompt template containing the user's history, conversation context, and journal prompt to request responses from an LLM (see supplementary materials for details on the prompt). After receiving the LLM-generated output, NLG calculates the sentence similarity between the output and the original journal prompt~\cite{uzun2022sentiment}.
This process filters out instances where the LLM hallucinates or generates irrelevant responses. If the sentence similarity score between the LLM output and the journal prompt is below 70\%, the Response Generation Module discards the LLM output and defaults to the original journal prompt.
The Response Generation Module also discards the LLM output if the generation takes more than $3ms$. We utilize OpenAI's GPT-3~\cite{brown2020language}, which was the most capable model available at the time of development. After personalizing the output from the LLM, the Response Generation Module then provides the journal prompt to the user.

%% file: new_sections/4-experiment_1.tex
\section{User Study}
This section details the design and execution of the user study, followed by data analysis and findings. 
The participant recruitment and study procedures were reviewed and approved by our Institutional Review Board (IRB) prior to the study, and only individuals included in the IRB had access to the study data.

\subsection{Study Design, Participants, and Execution}\label{sec:study_I_Design}

\subsubsection{Design}
\begin{figure*}[t]
    \centering
    \includegraphics[width=\linewidth]{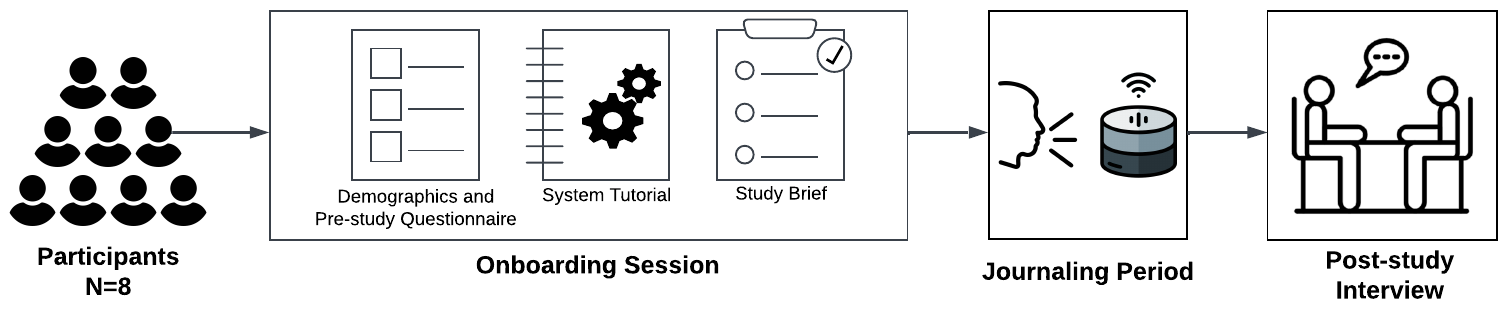}
    \caption{This figure illustrates the study workflow. After meeting participants in person and obtaining their signed consent, the study proceeded in three stages: (1) in-person onboarding, during which the participants completed a pre-study questionnaire to gather demographic information and their experience with journaling and smart devices. Then they watched a 12-minute tutorial on \toolname{}'s features and journaling examples, followed by a study brief with instructions on how to interact with \toolname{}, received a tutorial; (2) a two-week period where they used \toolname{} in their primary place of residence to journal their symptoms; and (3) a post-study interview where we collected their feedback on the conversation flow, session length, satisfaction, enjoyment, and views on personalization using conversation history.}
    \label{fig:study_design}
\end{figure*}

The study was designed in three stages (\autoref{fig:study_design}): onboarding, a two-week journaling period, and post-study interview. The in-person onboarding session introduced \toolname{} and gathered the participant's demographic and information related to their Parkinson's symptoms. During the two-week journaling period, participants used \toolname{} to document their symptoms, daily activities (e.g., exercise, meals, social interactions), and challenges. Then, we conducted a semi-structured interview to gather their feedback on their experience with \toolname{}.

\subsubsection{Pilot}
Prior to the main study, we conducted a pilot with a PwPD (not participating in the main study) to evaluate the study design and procedures, including onboarding instructions, test \toolname{}'s functionality, and ensure a two-week journaling period would suffice for collecting adequate data for meaningful analysis. The participant, recruited with the help of our team's neurologist, used \toolname{} on an Alexa Echo Dot (fourth generation) at least three times daily for two weeks. Based on pilot results, we refined onboarding instructions for clarity, addressed technical glitches, such as intent identification, and confirmed that a two-week journaling period provided adequate data by assessing the quality and quantity of the journal entries.

\subsubsection{Participants}\label{sec:rec-process}
We recruited participants with the assistance of our team's neurologist, who introduced the study to their cohort of Parkinson's patients. The neurologist also connected us with two local Parkinson's support groups, where the group leaders emailed their members about our study.
The lead author made several trips to the teaching hospital and nearby coffee shops, meeting interested patients in person to provide detailed explanations of the study's objectives. These efforts successfully recruited ten participants (five female, five male), all of whom were proficient in English\footnote{Given that \toolname{} operates in English, proficiency in the language was necessary to facilitate clear communication and avoid potential transcription errors during the conversational journaling process.} and clinically diagnosed with Parkinson's. Participation required completing the consent process by signing a consent form. One participant withdrew due to medical reasons before the study began, and another left a few days into the study for personal reasons. The eight participants who completed the study received a \$100 Amazon e-gift card and are referred to as [P$\#$].

\subsubsection{Execution}
We deployed \toolname{} in a two-week user study in which the participants used \toolname{} in their primary place of residence to document their symptoms, daily experiences, and challenges associated with their condition. The following sections provide more details about the study execution:

\pheading{In-person onboarding session:}
The lead author conducted in-person onboarding sessions with participants in public locations near their residences. During these sessions, we collected demographic information and had participants complete a pre-study questionnaire on their journaling and smart device experience (see supplementary materials). Each participant received an Alexa Dot fourth generation with \toolname{} preinstalled. A 12-minute tutorial demonstrated \toolname{}'s features and limitations, followed by journaling examples to illustrate how the system works. Participants then tested \toolname{} for $\approx$30 minutes and familiarized themselves with the prototype. 
We also gathered brief information about participants' Parkinson's symptoms, medications, and the daily challenges they face while managing the condition.
Finally, participants were instructed to use \toolname{} at least three times daily for two weeks.

\pheading{Journaling period}
During the two-week journaling period, participants used \toolname{} in their primary place of residence to routinely record their symptoms. The lead author conducted regular email check-ins with each participant every two days to ensure they were not facing any technical difficulties with \toolname{} and to provide assistance as needed. Journal data was encrypted and stored on a HIPAA-compliant AWS server, only accessible to the research team.

\pheading{Post-study interview}
At the end of the two-week journaling period, we conducted semi-structured post-study interviews with each participant to gather feedback on the conversation flow, session length, satisfaction, enjoyment, views on personalization and the usability of \toolname{}.
To streamline the interview process for our participants, the first author shared their screen, displaying statements with 5-scale Likert options (e.g., ``The system instructions were easy to understand and follow'': Strongly agree, agree, neutral, disagree, strongly disagree).
The interviewee would select an option followed by an open-ended discussion to elaborate on their answers on each of the above mentioned questions. We also asked several open-ended questions, such as their journaling practice and suggestions for improving \toolname{} (to see a more detailed list of questions, see supplementary materials). 
Each interview lasted 15-20 minutes via Zoom, and detailed notes were taken for later analysis.

\subsection{Data Analysis and Findings}\label{sec:data-analysis-study-1}
We began our data analysis by manually reviewing all collected journaling sessions. We removed ``null journals,'' where participants started a session but didn't proceed to journal, and merged consecutive sessions if a participant experienced a disruption and restarted within 30 seconds of ending the previous session. After cleaning the data, we closely examined the conversations between participants and \toolname{} to ensure data quality for reliable analysis. We quickly identified issues with intent identification, follow-up questions, and personalization. In 25\% of cases (24/92), \toolname{} either failed to identify or misidentified intents, leading to irrelevant or missing follow-up questions and halting conversation progress. Additionally, only 31\% of system responses\footnote{System response refers to all spoken output generated by the system, including follow-up questions and prompts. When referring to each category individually, we use `follow-up' or `prompts.' } were properly personalized with participant-specific details like names and medications. To address these issues, we conducted an in-depth analysis of the journaling data, system logs, and participant feedback.

\subsubsection{Intent identification issues}
A deeper investigation revealed that \toolname{}'s sub-optimal intent identification performance stemmed from three primary factors: 1) a \textit{``bulk journaling''} behavior, 2) voice-to-text transcription errors, and 3) inadequate performance of ``DIETClassifier''.

\pheading{Bulk journaling:} Our analysis revealed a previously unreported ``bulk journaling'' behavior in which participants' (five out of eight) entries were comprised of long, fragmented, and complex statements containing multiple topics from daily activities to symptoms and emotional experiences. For instance, P1 reported,  \textit{``I've been pretty active today doing strength training, stationary biking, and stretching exercises. Later on, I'm going swimming, and after that, I'm going dancing. I haven't noticed any symptoms today, and I took my medications on time.''} This ``bulk journaling'' behavior contrasted with our expected ``atomic journaling,'' where participants would articulate issues using concise, single-topic sentences across multiple exchanges with the system. Our analysis identified 37 instances of bulk journaling among five of the eight participants. Bulk journaling disrupted intent identification because the NLU was not trained to handle multiple intents in a single statement. As a result, the system either focused on a single issue, missing others or became unresponsive.

\pheading{Voice-to-text transcription errors:}
Transcription errors (19 instances) also hindered intent identification. For example, ``\textit{Carbidopa-Levodopa}'' was incorrectly transcribed as \textit{``...I take carpet leave a dopa...''}. While it is unclear what caused these errors post-study, possible factors include Alexa's transcription mechanism errors, voice fluctuations experienced by some PwPD, or background noise interference.

\pheading{DIETClassifier performance issues:} 
The DIETClassifier model's limited generalizability, compared to more flexible LLMs, caused the system to struggle with symptoms not covered in our training data (e.g., fatigue), leading to 50 instances of conversational breakdowns. This model often failed to handle unexpected user responses, making it difficult to steer the conversation back on track. Intent identification errors subsequently impacted \toolname{}'s ability to generate relevant follow-up questions. For instance, when P2 mentioned ``fatigue,'' the system incorrectly responded with, ``\textit{Can I ask you a few questions about your tremor?}'' A manual review of 1,266 follow-up questions revealed 200 instances (16\%) of irrelevant prompts.

\subsubsection{Personalization issues}
We manually examined and coded all questions and content generated by \toolname{} to investigate suboptimal personalization. This analysis assessed the system's use of participants' names and its ability to dynamically adapt questions based on their profiles and journaling history. Results showed that only 31\% of the follow-up questions were personalized.

Given the issues identified, we determined that the collected data might compromise our study's validity. Thus, we opted to update \toolname{} and conduct a second user study. 
However, we continued analyzing participants' post-study feedback.
Here, we present their qualitative responses, while for the Likert scale questions, we report only their levels of engagement and satisfaction.

\begin{table*}[t]
    \centering
    \caption{This table lists the six Parkinson's symptoms that were added to the system based on study findings, along with the corresponding follow-up topics and example questions. The color-coding shows the relationship between probing topics and symptoms. Upon detecting a symptom in a patient's utterance, \toolname{} uses the probing topics to ask follow-up questions.}
    \includegraphics[width=\linewidth]{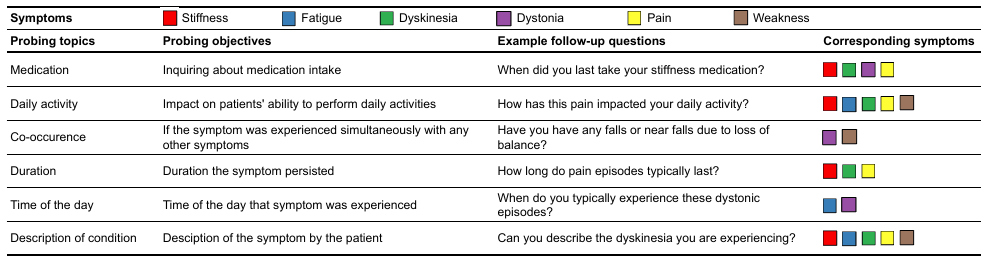}
    \label{tab:sympthoms_and_topics_expt2}
\end{table*}

\subsubsection{Analysis of post-study interview} We analyzed interview data, aiming to learn about participants' experiences and feedback on journaling with \toolname{}.

\pheading{Inadequate awareness of system's status:} 
Our analysis revealed that some participants abandoned or restarted journaling sessions due to being unaware of the system's busy status while processing responses. Although we explained during onboarding that Alexa's blue light indicates it is preparing a response, this visual cue may not have been sufficient for all participants. Some mentioned the ``silence'' during these periods confused them. For example, P4 said, \textit{``It would be nice if it had more feedback... there would be silence after I said I have tremors.''} Similarly, P6 found the pauses ``\textit{kind of annoying},'' and P8 noted that the lack of feedback made it difficult to understand the system. These comments suggest adding audio feedback could help users better recognize the system's status.

\pheading {Personalization:}
Participants expressed a positive view of personalization in \toolname{}. Four participants were pleasantly surprised by the personalized prompts from \toolname{}. For instance, P2 mentioned, \textit{``It kind of surprised me that it [\toolname{}] knew what medication I was taking.''} Personalization through conversation history also helped build rapport with the system. P1 and P6 noted, respectively, \textit{``The initial conversations were simple, but its [\toolname{}'s] understanding seemed to evolve over time,''} and \textit{``I would see it [\toolname{}] getting better as I used it more and got more comfortable with it.''} Additionally, P6 found that personalization motivated them to engage longer: \textit{``It was interesting to see such a [personalized] response, and most of the time, it made me continue the chat longer.''}

\pheading{Engagement level:}
Most participants found their interactions with \toolname{} engaging (three$\rightarrow$very engaging, three$\rightarrow$engaging, one$\rightarrow$neutral). P4 commented, \textit{``It [\toolname{}] was very easy to use, friendly, and handy.''} P2 noted, \textit{``It was fun [to converse with \toolname{}].''} However, P8 felt the conversation was ``non-engaging'' and mentioned during the post-study interview, \textit{``I would need much more direction [to use it properly].''}

\pheading{Satisfaction:} 
Overall, participants were satisfied with their journaling experience with \toolname{} (five$\rightarrow$satisfied, three$\rightarrow$neutral). P6 noted, \textit{``Got the hang of it very quickly, and it went pretty smoothly.''} However, two participants preferred more casual interactions. P5 mentioned, \textit{``I expected [to have] more casual conversations with it,''} while P2 said, \textit{``[The conversation] didn't go as expected as with a human.''}

%% file: new_sections/5-system-improvements.tex
\section{System improvements}
To move forward, we made several changes in \toolname{} to address issues identified in the user study (Study I hereafter).

\subsection{Expanding the list of symptoms}
Our analysis identified six additional symptoms that were not supported by \toolname{}, and the participants expressed a desire to record. In particular, 50\% of participants (four out of eight) wanted to record stiffness and fatigue, 37.5\% (three out of eight) wanted to document physical pain, and 12.5\% (one out of eight) wanted to record their experience with dyskinesia, dystonia, and weakness.
Based on these findings, we revised \toolname{}'s list of supported symptoms to include stiffness, fatigue, dyskinesia, dystonia, balance, pain, and weakness. 
We assigned probing questions to each of these newly added symptoms based on the recommendations of our neurologist and after extensive discussions with the entire research team.
Table~\ref{tab:sympthoms_and_topics_expt2} summarizes the newly added symptoms, their respective probing topics, and example follow-up questions.

\subsection{Improving intent identification}

To address intent identification challenges that emerged in Study I, we implemented an LLM-based intent interpreter.
Previous research in NLP has shown that LLMs are effective for conversational agents in identifying a wide range of intents~\cite{wei2024leveraging}.
To select an LLM, we ran experiments with three LLMs, GPT-4~\cite{achiam2023gpt}, Meditron~\cite{chen2023meditron}, and FlanT5~\cite{chung2024scaling}. Among the three LLMs, Meditron was specifically pre-trained to adapt to the medical domain.
We used FlanT5's pretrained XXL version and Meditron's 7B parameter models, both of which are available on Huggingface~\cite{wolf2019huggingface}.
To access GPT-4, we used OpenAI's API endpoint with the temperature set to $0$ and all other parameters set to default. To protect the patient's privacy, we paid careful attention that the conversation history shared with the API did not include personal identifiers such as the patient's name and age. As a complementary measure, we explored using the GPT-4 API's zero data retention (ZDR) feature to delete journaling data after each call immediately. However, OpenAI currently limits ZDR to projects covered by HIPAA regulations.

We used the data from Study I to tune the LLM's prompt and select the LLM with the best performance for intent identification.
To tune LLM's prompt for intent prediction, we used in-context learning (ICL)~\cite{chen-etal-2022-meta}, which included pairs of example user utterances and the corresponding intent in the prompt.
Each intent in the prompt corresponded to a symptom and included additional intents for audio speech recognition errors (``asr''), the user's refusal to discuss any symptom (``none''), and the user's expressed intent to discuss multiple symptoms (``multiple'').
We added the ``multiple'' intent based on our findings from Study I, which revealed that users frequently respond with a bulk of information covering multiple symptoms in a single utterance. 
When bulk journaling occurs, the intent interpreter classifies the utterance as a ``multiple'' intent. It then parses the user's response to identify each individual symptom intents. Each predicted symptom is subsequently treated as a separate intent and passed to the response generation module to prompt appropriate follow-up questions.
We ensured that the prompt design adhered to OpenAI's guidelines~\cite{promptengineering2024} and followed best practices from existing prompt-engineering research~\cite{kunnath2023prompting, feng2024prompting} (see supplementary materials for the prompt).
The tuned prompt was consistent across all three models.

To evaluate the three LLM models, the lead author manually reviewed the user-reported conditions, symptoms, and requests from Study I to code the ground truth intent.
Using coded ground truth data, we assessed the accuracy of each of the three models for intent identification. 
The accuracy was determined by dividing the number of correct predictions by the total number of predictions.
Our results showed that GPT-4 (accuracy=$95\%$) significantly outperformed FlanT5 (accuracy=$36\%$) and Meditron (accuracy=$1\%$) for intent identification(see supplementary materials for results).
Based on our findings, we used the prompt-tuned GPT-4 to build our LLM-based intent interpreter.

\begin{figure*}[t]
    \centering
    \includegraphics[width=\linewidth]{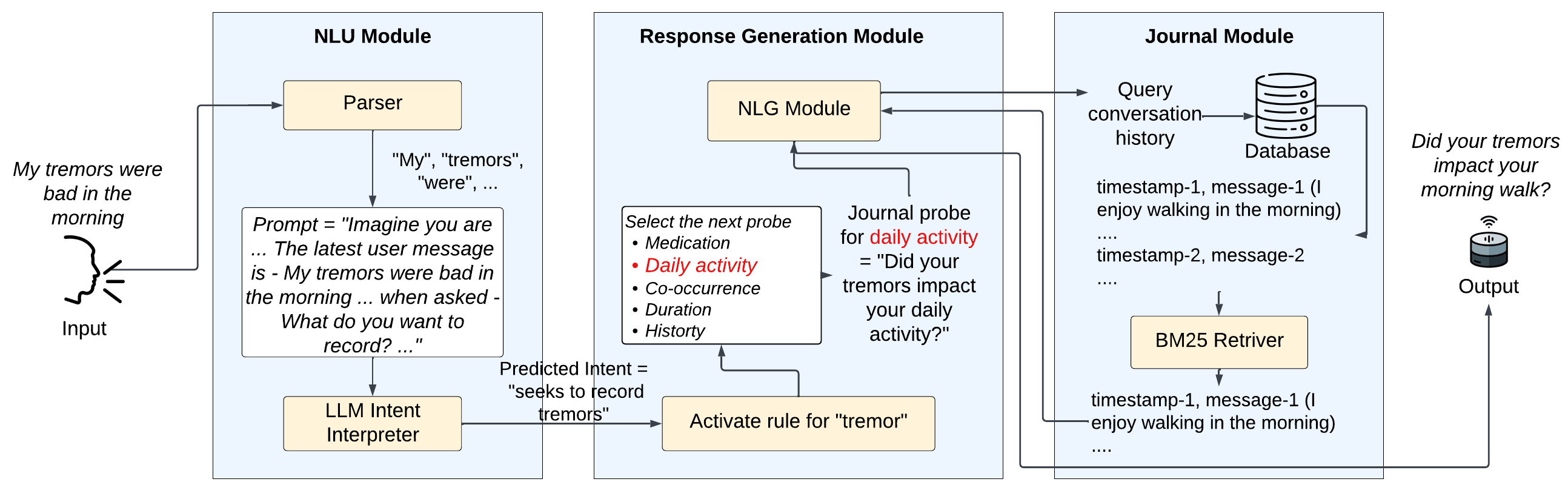}
    \caption{The system overview of \toolname{} following improvements made after Study I. First, the NLU module parses the user's input and forwards the information to the LLM intent interpreter as a prompt. The LLM intent interpreter predicts the user's intent from the prompt and sends it to the response generation module. The response generation module uses the predicted intent to trigger an appropriate journaling rule and selects the next relevant journal probe. To personalize this probe, the response generation module queries the journal module using the journal probe and conversation context. The journal module uses a BM25 retriever to retrieve relevant history from the user's conversation history that correspond to the journal probe and the ongoing conversation. This conversation history is then used to personalize the journal probe, which is sent to the user through Alexa.
}
    \label{fig:system_overview_expt2}
\end{figure*}

\subsection{Improving the accuracy of follow-up question generation}
We updated the response generation module to alleviate the issues of incorrect follow-up questions that occurred in Study I.
To this end, we used a rule-based classifier to generate a response for \toolname{}.
The rule-based classifier predicts a symptom based on the user's input and then asks follow-up questions designed to probe specific topics related to that symptom. 
Unlike the DIETClassifier used in Study I, which predicts a follow-up question based on the conversation context, implementing rules ensures that follow-up questions align with our intended probing topics and adhere to our conversation design considerations.

We implemented a rule for all 13 symptoms in our rule-based response generation module design. The rules are triggered and activated whenever a user mentions a symptom to \toolname{}. When a rule is activated, the response module generates follow-up questions based on the probing topics that correspond to the rule.
For example, saying, ``I have tremors in my left hand,'' triggers the rule that prompts follow-up questions about medication information, daily activity, and the duration of tremors.
After the user responds to \toolname{}'s follow-up questions, the NLU module computes their intent to determine whether they require clarification, want to skip the current question, or wish to end the journaling session. 
If the user's intent indicates that they want a clarification, the response generation module gives an appropriate answer and then asks the same follow-up questions.
We prompt GPT-4 with the conversation context and the user's query to generate an appropriate response.
Moreover, if the user intends to skip the current follow-up question, the response generation module moves to the next available follow-up questions.
Conversely, if the user wishes to end the journaling session, the response generation module acknowledges the request and ends the conversation by deactivating the rule.
Finally, if the user's response does not indicate a request for clarification, a desire to skip the current follow-up, or an intention to end the journaling session, the response generation module accepts the input and continues the conversation.
We upgraded the response generation module from GPT-3 to GPT-4 to utilize its 50\% higher input token limit, enabling more context-rich accurate responses~\cite{salemi2024optimization}. In addition, previous studies have shown that GPT-4 outperforms GPT-3 in generating contextually and medically accurate responses~\cite{rosol2023evaluation, stribling2024model, achiam2023gpt}.

The rules in our response generation module are mutually disjoint and can be nested within each other.
This could be beneficial for cases where the users want to switch topics during the conversation and then return to the previous ones.
For instance, if a user is discussing their tremors but wishes to talk about their mood, the response generation module activates the mood rule and nests it under the tremor rule.
By nesting the mood journaling rule after the tremor journaling rule, the user is able to journal their tremors after journaling about their mood.

\subsection{Improving personalization performance}

Analysis from Study I revealed that only $31\%$ of the outputs generated by the response generation module incorporated dialogue history, such as the participant's name and medication information. 
To improve this performance, we integrated a retrieval algorithm (see Figure~\ref{fig:system_overview_expt2}) that compares the current conversation context with the dialogue history to retrieve relevant data, which is added to the prompt of the LLM. This RAG method is shown to improve the performance of LLMs for generating personalized outputs~\cite{salemi2024optimization}.
We chose BM-25~\cite{robertson1995okapi} for our retrieval algorithm due to its low computational requirements and ability to scale to large amounts of data.
We evaluated the journal module with the retrieval algorithm using data from Study I and observed that $88\%$ of the response generation module's outputs included content from dialogue history, indicating a $66\%$ improvement. 
In addition to retrieving relevant dialogue history for the LLM prompt, the retrieval algorithm assisted the journal module in prioritizing relevant dialogue history and excluding data when the prompt exceeded GPT-4's token limit of $8192$.
 
\subsection{Incorporating audio feedback}
Several participants in Study I (six out of eight) reported confusion and frustration arising from not being aware of the system's status while it was preparing replies. Research~\cite{brauer2022alexa} in voice-enabled conversational systems has shown that audio feedback indicates that the system is processing user input and reduces confusion in conversation turn-taking. Therefore, we integrated an auditory feedback mechanism into the system, which played a ``pen-writing-on-paper'' sound file while participants awaited responses from the system. We chose the pen-on-paper-writing sound to indicate that the system was busy actively ``noting'' user inputs.

%% file: new_sections/6-experiment_2.tex
\section{Second User Study (Study II)}

We deployed \toolname{} in a two-week user study with nine PwPD, who used it to journal their daily challenges related to their condition. 

\subsection{Study Design, Participants, and Execution}
With the exception of the updates made to \toolname{} and minor adjustments to the onboarding session, the designs of Study I and Study II were identical (\autoref{fig:study_design}). Therefore, the design details of Study II are not repeated in this section.

\subsubsection{Participants}
We recruited nine PwPD (five female, four male), including six returning (four female, two male) and three new participants (one female, two male). We refer to returning and new participants as [RP$\#$] and [NP$\#$], respectively, using the same numerical index for returning participants as in Study I for comparison (e.g., RP1 in Study II is P1 in Study I). The three new participants were recruited using the same procedure detailed in~\ref{sec:rec-process}. Recruiting both returning and new participants allowed us to evaluate the system's improvements while gaining fresh perspectives for broader insights.

\subsubsection{Execution}
Similar to Study I, we deployed \toolname{} for two weeks, during which the nine participants used the system in their primary place of residence to document their daily experiences and challenges associated with their condition. The journaling period and post-study procedure were identical to Study I (\autoref{sec:study_I_Design}).

\subsection{Data Analysis and Findings}
We analyzed the 251 journaling sessions and participants' feedback to address our research questions. Below, we present our data analysis corresponding to each research question. Unless specified otherwise, three team members conducted qualitative analyses of all participants' utterances as well as system responses using the following methodology: they jointly coded three shared data subsets, compared notes to ensure consensus, and then independently worked on two additional subsets, each.

\begin{table*}
\centering
\setlength{\extrarowheight}{0pt}
\addtolength{\extrarowheight}{\aboverulesep}
\addtolength{\extrarowheight}{\belowrulesep}
\setlength{\aboverulesep}{0pt}
\setlength{\belowrulesep}{0pt}
\caption{This table shows the average intent identification accuracy, the relevance of system responses, the relevance of participants' responses, and the personalization rate for participants in Study II. The gray cells highlight the mean and standard deviation for each metric across all participants. In the table, RP\# refers to returning participants from Study I, while NP\# represents newly recruited participants for Study II.}
\label{tbl:combined_results}
\begin{tabular}{llllllllllll} 
\toprule
                                    & \textbf{RP1} & \textbf{RP2}  & \textbf{RP3}  & \textbf{RP4}  & \textbf{RP5}  & \textbf{RP6}  & \textbf{NP7}  & \textbf{NP8}  & \textbf{NP9}  & {\cellcolor[rgb]{0.929,0.929,0.929}}\textbf{Mean} & {\cellcolor[rgb]{0.929,0.929,0.929}}\textbf{SD}  \\ 
\midrule
Intent identification accuracy      & 0.97         & 0.98          & \textbf{0.99} & \textbf{0.99} & \textbf{0.99} & \textbf{0.99} & \textbf{0.99} & \textbf{0.99} & \textbf{0.99} & {\cellcolor[rgb]{0.929,0.929,0.929}}0.99          & {\cellcolor[rgb]{0.929,0.929,0.929}}0.01         \\
Relevance of system responses       & 0.98         & \textbf{0.99} & 0.98          & 0.96          & \textbf{0.99} & 0.95          & 0.91          & \textbf{0.99} & 0.97          & {\cellcolor[rgb]{0.929,0.929,0.929}}0.97          & {\cellcolor[rgb]{0.929,0.929,0.929}}0.03         \\ 
\hline
Relevance of participants responses & 0.95         & 0.98          & \textbf{0.99} & 0.99          & 0.94          & 0.88          & 0.95          & 0.94          & 0.98          & {\cellcolor[rgb]{0.929,0.929,0.929}}0.96          & {\cellcolor[rgb]{0.929,0.929,0.929}}0.04         \\ 
\hline
Personalized Responses              & 66           & 110           & 28            & 67            & 12            & 64            & 5             & 40            & 82            & {\cellcolor[rgb]{0.929,0.929,0.929}}–             & {\cellcolor[rgb]{0.929,0.929,0.929}}–            \\
Total responses                     & 82           & 155           & 34            & 71            & 15            & 74            & 7             & 55            & 92            & {\cellcolor[rgb]{0.929,0.929,0.929}}–             & {\cellcolor[rgb]{0.929,0.929,0.929}}–            \\
Personalization rate                & 0.8          & 0.71          & 0.82          & \textbf{0.94} & 0.8           & 0.86          & 0.71          & 0.73          & 0.89          & {\cellcolor[rgb]{0.929,0.929,0.929}}0.81          & {\cellcolor[rgb]{0.929,0.929,0.929}}0.08         \\
\bottomrule
\end{tabular}
\end{table*}

\subsubsection{How can we effectively simulate the clinical interview process through targeted, relevant, and personalized questions?}
\hspace{7pt}

An essential part of accurately simulating the clinical interview process is correctly identifying the symptoms the patient reports (i.e., intent). Therefore, we began by assessing the accuracy of intent identification in Study II,  using system responses to participant inputs as a proxy. For instance, following the input  \textit{``I want to record my \underline{tremors}''} with \textit{``Did your \underline{tremors} impact your daily activity?''} indicates correctly identifying the participant's intent to record information about their tremor. Conversely, following the same input with \textit{``When did you last take your \underline{pain} medication?''} indicates inaccurate intent identification. We used binary coding (0$\rightarrow$inaccurate, 1$\rightarrow$accurate) to categorize the results. Our analysis revealed an average intent identification accuracy of 99\% (Std:0.01\%, Range [97\%, 99\%]), a 29\% improvement from Study I's 70\% accuracy.  We also evaluated the relevance of system responses based on the identified intent and the participant's profile (0$\rightarrow$non-relevant, 1$\rightarrow$relevant). An example of an irrelevant response would be asking a participant who reports sleeplessness but does not take any known medications for it about their sleep medication adherence. Across all participants, we identified 38 non-relevant system responses (Avg. 97\%, Std: 0.03\%, Range [91\%, 99\%]), 20 of which seemed triggered by incomplete participant inputs. For instance, RP2's incomplete utterance, \textit{``I would like to record successful,''} prompted the system to respond with \textit{``Please go ahead and tell me about the successful event you'd like to record,''} followed immediately by a non-relevant prompt \textit{``I have successfully recorded your response. Please say ``exit'' to close the conversation or say ``restart'' to start a new conversation.''} Table~\ref{tbl:combined_results} presents the accuracy and relevance breakdown per participant.

To assess \toolname{}'s personalization performance, we manually coded (0$\rightarrow$not correctly personalized, 1$\rightarrow$correctly personalized) system responses. ``Correct personalization'' involved dynamic adaptation based on each participant's profile and journaling history, which included their name, medications, prevalent symptoms, and past interactions. Of the 585 responses generated by \toolname{}, 474 were correctly personalized. For example, when NP9 reported \textit{``I'm continuing to have stiffness in my left hand which curls up my fingers,''} the system correctly personalized its response, \textit{``When was your last Carbidopa-Levodopa dose?''} by retrieving the medication taken for stiffness by NP9 from their profile. We found an overall personalization accuracy of 81\% (std: 8\%, Range [71\%, 94\%]) across all participants, a significant improvement from the 31\% in Study I. We also calculated the personalization rate for each participant by dividing correctly personalized instances by the total system responses (\autoref{tbl:combined_results}).

The combination of high intent identification accuracy, relevant responses, and effective personalization demonstrates the feasibility of creating conversational journaling systems that successfully engage participants in meaningful and coherent health-related dialogues.

\subsubsection{RQ2: How can we solicit responses from PwPD that facilitate rich data collection and offer clinical insights, deepening understanding of patient experiences and disease management?}
\hspace{7pt}

Collecting clinically valuable data through conversations hinges on patients providing relevant responses to the follow-up questions. We analyzed and coded the relevance of participants' responses to the system's inquiries, considering responses relevant if they directly addressed the question and offered specific, logical information about the patient's condition. For instance, a non-relevant response was RP6 answering, \textit{``When did you last take your medication for stiffness?''} with \textit{``no''} [perhaps misunderstanding the question as ``did you take you stiffness medication?'']. Our analysis found that 96\% (Std: 0.4\%, Range [88\%, 99\%]) of responses were relevant, with only minor differences across participants (Table~\ref{tbl:combined_results}).

\begin{figure*}[t]
    \centering
    \includegraphics[width=\linewidth]{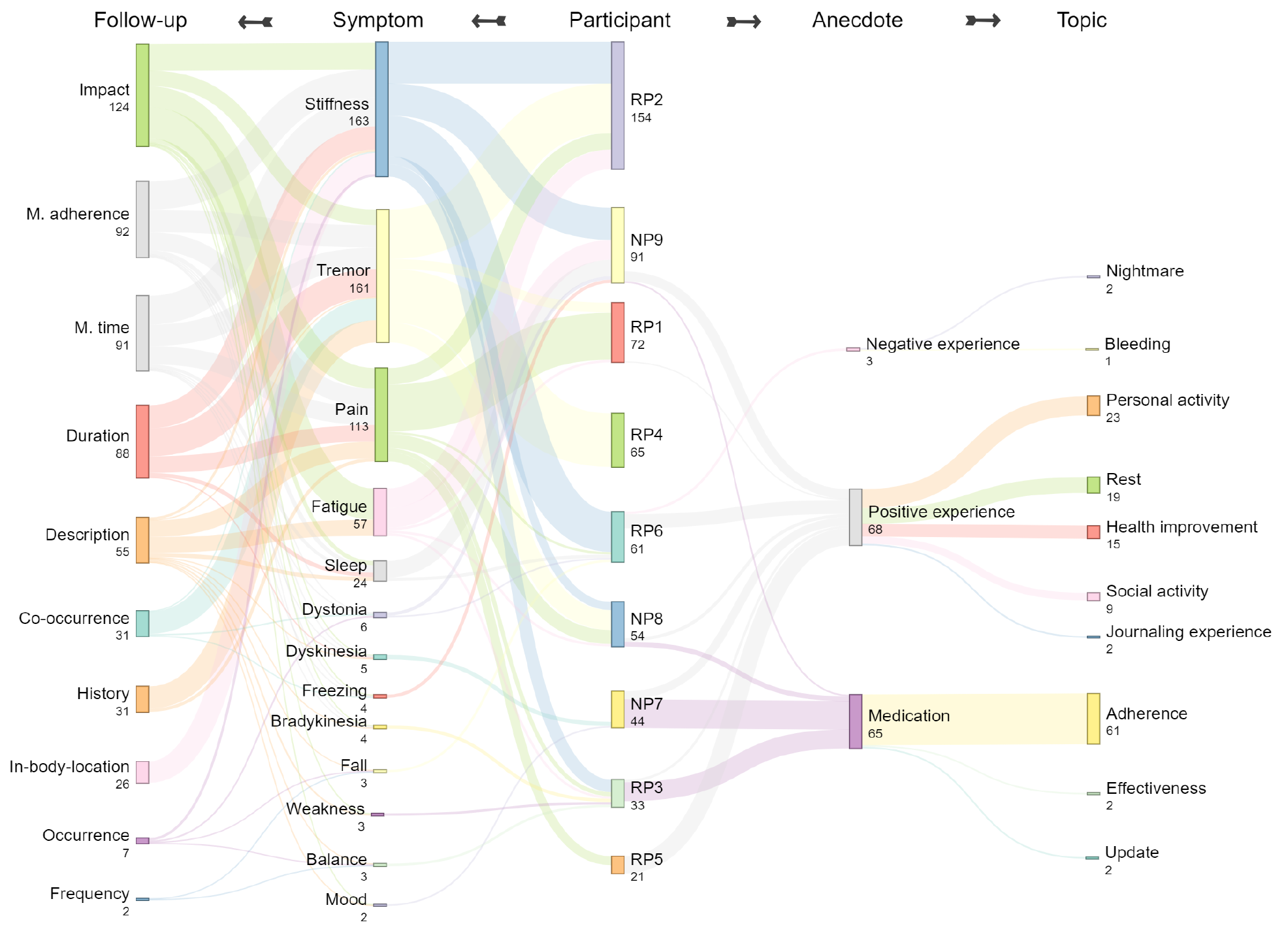}
    \caption{This diagram shows information captured via journaling in Study II. To the left of Participant(s) are Symptom(s) expressed and the Follow-up(s) generated by \toolname{} to further probe and collect information about each symptom.  To the right of Participant(s) are Anecdote(s) and Topic(s).  Anecdotes are expressions that are not directly related to one of the 12 symptoms included in our study but provide valuable insights about participants, such as information about their medication and positive or negative experiences. Participants are ordered in descending order by the total number of symptoms and anecdotes.}
    \label{fig:sankey}
\end{figure*}

\begin{figure*}[t]
    \centering
    \includegraphics[width=0.9\linewidth]{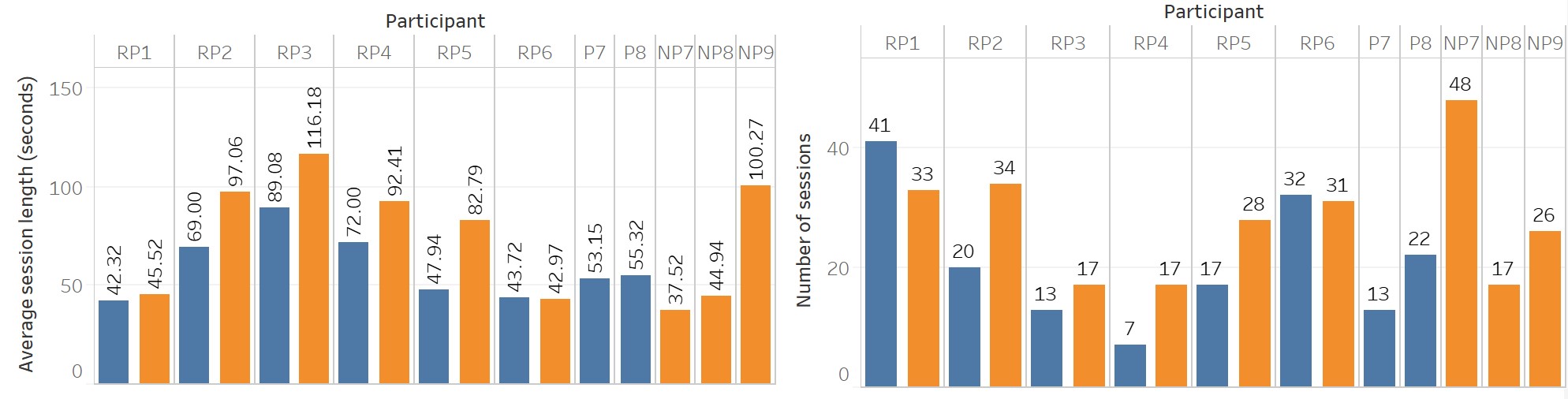}
    \caption{The two graphs show the average journaling session lengths (left) and the total number of sessions (right) for participants in Study I (\includegraphics[width=1.5ex]{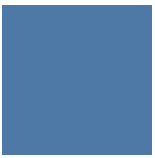}) and Study II (\includegraphics[width=1.5ex]{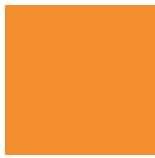}). Each participant is identified as either a returning participant from Study I (RP\#) or a new participant in Study II (NP\#). Participants P7 and P8 were in Study I but did not return for Study II. The left chart shows the average time spent per session in seconds, while the right chart shows the total number of journaling sessions completed by each participant in both studies.}
    \label{fig:number_of_sessions}
\end{figure*}

To assess the informativeness of the collected data, we qualitatively analyzed all the participants' responses, checking if they directly (e.g., tremor) or indirectly (e.g., shaking) referenced any of the 12 Parkinson's symptoms in our study. Capturing each reported symptom can help caregivers of PwPD better understand their experiences between clinical visits. For each symptom, we then examined participants' responses to the follow-up questions, which provided a richer, more nuanced view of their experiences. During the analysis, we identified utterances not directly related to a symptom but still valuable for understanding participants' experiences and behavior, which we termed as ``anecdotes.'' We classified anecdotes into three types: ``negative experiences,'' ``positive experiences,'' and ``medication.'' Positive experiences (e.g., RP1: \textit{``I feel good today''}), negative experiences (e.g., RP6: \textit{``I had a nightmare about falling''}), and medication anecdotes involve information about medication adherence and updates. Unlike symptom-specific follow-up questions about medication generated by the system, medication anecdotes were initiated by the participants. Examples include NP7, \textit{``at 1220 I took a carbidopa levodopa''} and NP8, \textit{``I'm changing the frequency of my medication, and it seems to be working a little bit better.''} Figure~\ref{fig:sankey} shows symptoms expressed by participants, system-generated follow-ups as well as anecdotes, and their topics. The high relevance of responses, combined with expressed symptoms and follow-up data, demonstrates the feasibility of collecting rich, multifaceted data via conversation journaling. For example, during two weeks of journaling, RP2 reported stiffness, tremor, pain, and fatigue, with follow-ups exploring their \textit{impact} on daily activities, \textit{medication adherence} (M. adherence), \textit{time} (M. time), and more.

To understand factors that could potentially influence journaling behavior, we examined the relationship between the number of years diagnosed with Parkinson's and the number of symptoms expressed by each participant. This was, however, inconclusive. For instance, RP2, who recorded the most symptoms (154), and RP5, who recorded the fewest (21), had both been living with Parkinson's for nearly nine years. We also compared the average length and number of journaling sessions between the six PwPD who participated in both studies (\autoref{fig:number_of_sessions}). The journaling behaviour between and within participants varied with no strong recurring patterns, suggesting that journaling behavior might be circumstantial. Personal circumstances might have influenced journaling behavior. For instance, RP4 noted that \textit{``most of the time my answer was slight tremor in right arm cause that's the only real symptom I had.''} Other circumstances, like being away or busy, also affected journaling frequency. For example, RP2 mentioned, \textit{``I did it three times a day… [but] it may have been one or two days while I was away in the afternoon.''} and RP1 said \textit{``you get busy in your life and you forget [to journal].''}

To evaluate the clinical value of the data, we had our team's neurologist independently review all journal entries and provide feedback on their potential for patient care. They highlighted the richness of the information, noting, \textit{``it is impressive how much detailed information is there,''} and emphasized its value in extending their understanding of patients' daily experiences, stating, \textit{``from this data, one can begin to appreciate the range of symptoms occurring throughout the typical daily experience in PwPD.''}
We also sought feedback from two independent movement disorder specialists who were not involved in the research. We provided them with a detailed description of \toolname{}, the study procedure, goals, and anonymized participant data. The first expert emphasized the value of conversational journaling for capturing crucial information often difficult to obtain during clinical visits:
\begin{quote}
    \textit{``Its main advantage is that it is voice-activated and asks the patient's follow-up questions. This information would be crucial in a clinical encounter as it can highlight the temporal pattern of motor and non-motor fluctuations, physical activities, and mental states. This information is commonly challenging to obtain in clinics; Patrika offers a potential solution.''} 
\end{quote}

After reviewing the data collected by \toolname{}, the second expert highlighted the advantages of conversational journaling over traditional handwritten diaries for Parkinson's and its potential to enhance medical data collection and patient care:

\begin{quote}
    \textit{``Patients with Parkinson's have difficulty consistently using written diaries to track their symptoms... 
    Patrika enables patients to quickly record their symptoms and medication dosing, and Patrika's prompting allows for clarification. This digitally collected data has the potential to direct patient care.  With additional development, Patrika could be used to synthesize data into clinically meaningful narratives for the patient and their provider.''}
\end{quote}

\subsubsection{RQ3: How do PwPD evaluate the usefulness and effectiveness of conversational journaling for tracking and managing their condition?}
\hspace{7pt}
We analyzed participants' exit interview feedback, including responses to Likert scale questions and the subsequent open-ended discussions, to understand their experiences with \toolname{} and their perceived value of voice-enabled conversational journaling.

\begin{figure*}[t]
    \centering
    \includegraphics[width=0.85\linewidth]{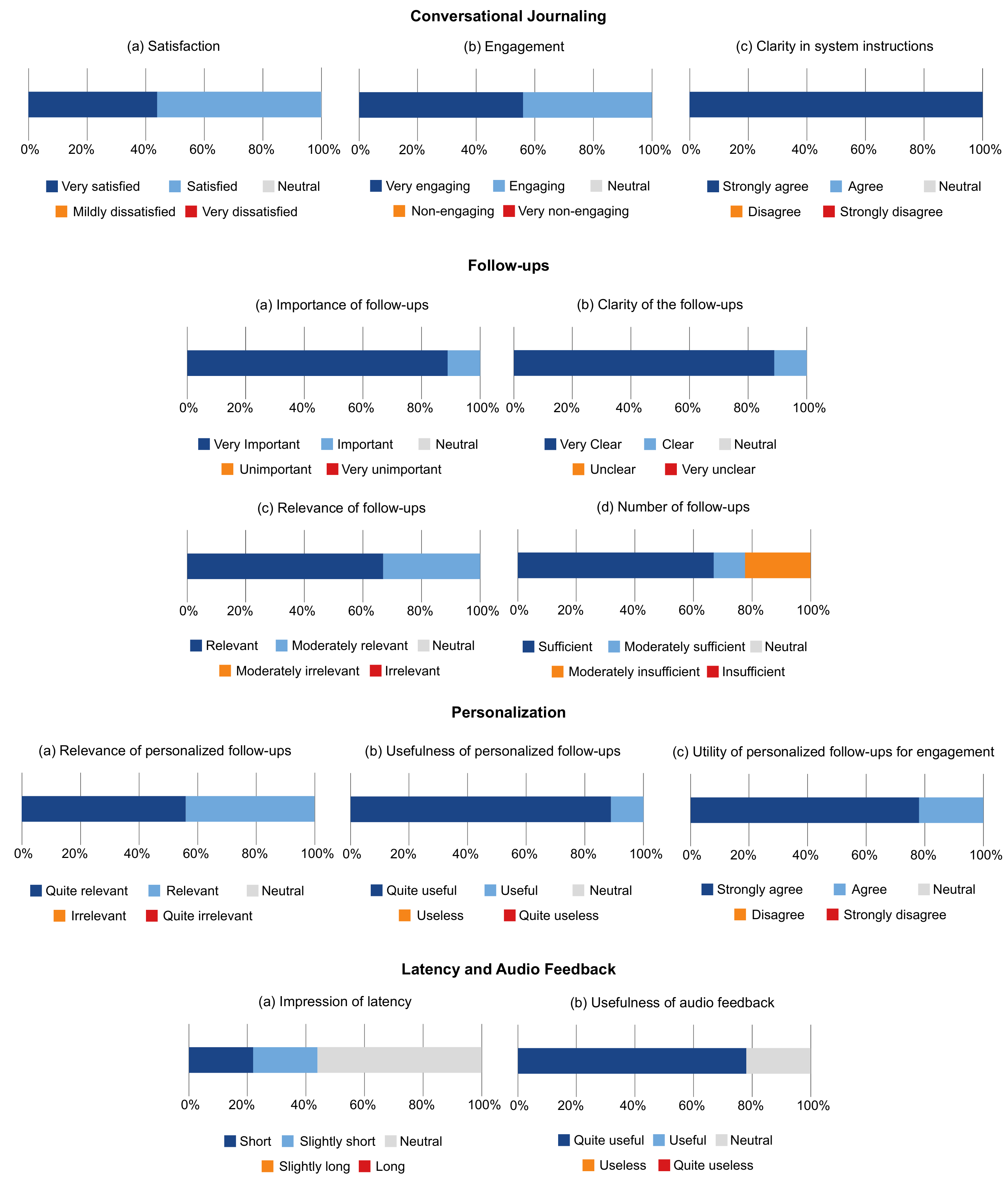}
    \caption{The figure shows participants' subjective assessment of \textbf{conversational journaling} experience, quality of \textbf{follow-ups} generated by the system and its \textbf{personalization} performance, and their impression of \textbf{latency} and the value of \textbf{audio feedback}. We use a single diverging categorical colormap to encode positive $\leftrightarrow$ negative responses for all items.}
    \label{tbl:analysis-combined}
\end{figure*}

\pheading{Conversational journaling:} 
We asked participants to rate their overall satisfaction with journaling using \toolname{} and whether they found conversational journaling engaging (\autoref{tbl:analysis-combined}). Responses were positive, with all participants expressing satisfaction (4$\rightarrow$very satisfied, 5$\rightarrow$satisfied) and finding conversational journaling engaging (5$\rightarrow$very engaging, 4$\rightarrow$engaging). Four participants (RP4, RP6, NP8, NP9) noted that the conversational aspect motivated them to journal. For instance, RP6 appreciated the interactive nature, stating, \textit{``having a conversation really, as opposed to just... a recording system''} while RP4 noted, \textit{``you talk back and forth to it… as opposed to…doing a web survey or something like clicking on it I mean. It was very engaging.''} NP7 and NP8 mentioned that clear enunciation was necessary but did not detract from their positive experience. NP7 noted, \textit{``I learned I have to enunciate very carefully …, but it's good practice.''} RP4 mentioned the lack of immediate access to journaling data as a drawback. This is, however, a limitation of the current design of \toolname{} rather than an inherent limitation of conversational journaling. All participants strongly agreed (9$\rightarrow$strongly agree) that the system's instructions were clear and easy to follow.

The incorporation of empathy in system responses was also noted and appreciated by RP3 and RP6. RP3 mentioned that the system's responses, such as, \textit{``I'm sorry you're feeling that way,''} conveyed empathy, further emphasizing, \textit{``So that little bit of compassion, even though it's coming from a machine, is nice.''} RP6 also noted, \textit{``When I [expressed I was] having a good night's sleep…it said, I'm glad you had a good night's sleep, and I was like, hey, that's pretty cool.''} This empathic reaction appears to have enhanced their overall experience with the system.

\pheading{Follow-ups:} 
We asked participants' opinions about the value of follow-up questions in enhancing data collection during journaling and their effectiveness within \toolname{} (\autoref{tbl:analysis-combined}). Participants strongly agreed on the importance of follow-ups for data enhancement (8$\rightarrow$very important, 1$\rightarrow$important) and found them clear (8$\rightarrow$very clear, 1$\rightarrow$clear) and appropriately timed (8$\rightarrow$just fine, 1$\rightarrow$short). Opinions on the relevance (6$\rightarrow$relevant, 3$\rightarrow$moderately relevant) and the number of follow-ups were more divided (6$\rightarrow$sufficient, 2$\rightarrow$moderately sufficient, 1$\rightarrow$moderately insufficient).

We further examined participants' open-ended comments to learn the reasons for these ratings. RP2, RP3, and RP4 emphasized the value of follow-ups to help clarify and delve deeper into the participants' responses. For instance, RP2 mentioned,\textit{ ``I think I probably like the most [about conversational journaling] is that it was asking me those questions,''} highlighting the value of the system’s inquiries. RP4 emphasized the importance of obtaining more detail from the system about their answers, stating, \textit{``It's very important … to get more detail [by the system] about the answer [that participant provided].''} Similarly, RP3 felt follow-up questions helped clarify their statements, ensuring the system understood their input: \textit{``It helped to clarify what I had stated and then I knew that the system really understood my question and my comment.'' } Interestingly, RP2 and RP3 also valued the follow-ups as reminders, with RP2 noting \textit{``It was good that they asked when I had taken my last medication, because sometimes I do forget and it was kind of a reminder.''} RP3 echoed this sentiment, noting that, \textit{``[questions functioned as] Reminder about when you took it or when you're next going to take it.'' } Opinions on the number of follow-up questions were divided. NP9, RP6, and NP8 wanted more follow-ups to enrich data collection. NP8 mentioned, \textit{``I thought there could be more, like more detailed questions.''}  RP6 suggested incorporating follow-ups for anecdotes: \textit{``I can come back from a bike ride and say I had a great bike ride, and it will say to me, oh, could you tell me more detail?''}  In contrast, RP3 and NP8 found the number and depth of follow-ups adequate, with RP3 stating, \textit{``They were just the right amount... I don't really need to get into a long conversation.''}

\pheading {Personalization:} We asked participants to share their opinions about the personalization of system responses (~\autoref{tbl:analysis-combined}). All participants found the references to prior conversations, the use of their names, and the system's awareness of their medical profile and medications to be relevant (5$\rightarrow$quite relevant, 4$\rightarrow$relevant) and quite useful (8$\rightarrow$quite useful, 1$\rightarrow$useful). Additionally, participants felt that personalization made journaling more engaging and relatable (7$\rightarrow$strongly agree, 2$\rightarrow$agree).

We examined participants' open-ended feedback to understand the reasons behind their positive assessments of personalization and identify possible pitfalls and challenges. Seven participants (RP1, RP2, RP3, RP6, NP7, NP8, NP9) felt personalization made the system seem more sensitive, caring, and attentive. RP2 appreciated that, \textit{``[it appeared as if] somebody was interested in how you were feeling,''} a sentiment echoed by RP1: \textit{``You know, it's like she heard you.''} NP7 and NP8 noted that the system made them feel cared for, with NP7 saying it felt like \textit{``somebody was caring enough to ask questions about what I was feeling,''} and NP8 adding, \textit{``It's like if I had a conversation with you, you would remember the conversation from yesterday.''} RP2 also valued the system using their name, which added a personal touch. Personalization of questions based on the user's profile was appreciated, with RP6 noting the system didn't ask unnecessary follow-ups: \textit{``If I told her I had a good sleep, it didn't ask me what sort of medicine I took.''} Interestingly, personalization helped RP3 recognize recurring issues by asking about past experiences. They appreciated this, noting, \textit{``I think it's good that it's brought in [the past experience], and then it makes the user aware that maybe there's a trend that looks it requires more attention.''}

RP1 noted that personalization may occasionally lead to the generation of imprecise follow-ups, such as inquiring about the impact of a symptom on daily activities they did not perform:~\textit{``she would ask the same questions… [about] my activities. I only gave her two activities...but I don't do them every day.''}  

\pheading{Latency and audio feedback:} 
 We asked participants to rate system latency and the utility of audio feedback in understanding the system's busy status (\autoref{tbl:analysis-combined}). Impressions of latency were mixed (2$\rightarrow$short, 2$\rightarrow$slightly short, 5$\rightarrow$neutral), but no one found the wait time frustrating. Most participants found the audio feedback very useful (7$\rightarrow$quite useful, 2$\rightarrow$neutral), especially returning participants who appreciated the feature. RP6 noted, \textit{``That [audio feedback]... was a huge thing to me because... when I heard the pencil scratching, I was like, OK, it's thinking right now,''} and RP5 mentioned, \textit{``I thought that was an excellent way of letting me… know that the computer had received what I was saying and was writing it down.''} RP1 and NP7 were neutral, with RP1 believing regular use would make audio feedback unnecessary.

\pheading{Voice as input modality:}
Six participants (RP1, RP3, RP6, NP7, NP8, NP9) highlighted the benefits of using voice as an input modality, praising its ease of access and efficiency. They appreciated the convenience of voice input, which removed the need for typing. RP6 noted, \textit{``Definitely better than other journaling systems I've used... I didn't have to write anything.''} RP1 echoed this, stating, \textit{``It's easier to talk and say what your symptoms are because you get lazy typing it all in.''} NP7 and NP8 also found it easy to use, with NP7 saying, \textit{I didn't have to write it all down. It was easy to use.''} and NP9 adding, \textit{``I like the ease of using it... I can just walk into the room and tell it whatever I need to tell it.''}

\pheading{Suggestions for Enhancing Journaling:}
RP6 recommended enabling the system to learn new symptoms, and participants suggested allowing users to update profiles and medications on the fly. RP6 stated, \textit{``We could import our medicines ourselves on the fly because meds change sometimes,''} and RP1 wanted to add activities such as walking. Several participants suggested proactive journaling, a model in which the system initiates journaling via reminders. RP6 proposed, \textit{``maybe even a question like is now a good time for you to journal for today?''} and RP4 recommended, \textit{``The little device say, hey, it's time to journal… [or] like some kind of notification like an e-mail or a text or something,''} and NP8 suggested, \textit{``If I haven't journaled by, say 3:00 in the afternoon, it would do a popup message.''} RP2 advised the system could diagnose issues and suggest performance improvement tips, such as asking PwPD to \textit{``speak louder.''} Finally, participants emphasized the importance of accessing and analyzing journaling data, with RP4 saying, \textit{``It would be good to be able to go to a website and see the data that's been collected and see what things are,''} and NP7 adding,\textit{ ``It would be nice to learn ... how I felt when I took the different medication scenarios, which ones worked well for which type of activity.''} Implementing these suggestions would improve the system's adaptability, usability, and overall user experience.

%% file: new_sections/7-discussion.tex
\section{Discussion and Future Implications}
Nunes et al. identify the need for new human-centered self-care technologies that enable capturing and documenting a broad spectrum of PwPD experiences while delivering individualized and tailored user experiences~\cite{nunes2019agency}. Through iterative design, development, and rigorous empirical evaluation of \toolname{}, we show that generating personalized and context-aware probing questions based on each patient's profile and history enables the collection of rich qualitative insights into various aspects of PwPD lived experiences, while enhancing the journaling experience. 

Enhancing data collection and the journaling experience for PwPD offers multifaceted prospective benefits. First is the potential clinical gains. Conversing with the system, participants disclosed their daily negative and positive experiences. Probing questions by \toolname{} further encouraged them to elaborate, clarify, and provide context, enhancing the depth and breadth of journal data. The potential for better understanding patients' experiences and improving care was highlighted by the three experts who reviewed the journal entries. Another clinically advantageous aspect of conversational journaling is capturing subjective motor (e.g., imbalance) and non-motor (e.g., mood) experiences of PwPD which may not be obtained or diminished by tracking technologies such as wearables. Second is the potential for improved self-reflection and care. Conversational journaling alleviates the common pitfall of limiting the logging of patient experiences to a format (e.g., gait data collected via wearables) that may not always be the most reflective for that individual~\cite{vafeiadou2021self}. A study~\cite{vafeiadou2021self} with an online community of PwPD highlights the mismatch between patients' self-management data needs and what is currently supported by clinical monitoring tools. Finally, conversational journaling may decrease caregivers' burden if the patient is able to journal their experiences independently. Participants in our work generally found using \toolname{} to be a low-effort and engaging experience. Using voice as an interaction modality can lower the access barrier, enabling PwPD even with substantial motor or cognitive impairment to be able to journal.

Our work contributes to the field of HCI by advancing the understanding of human-centered and technical considerations for designing AI-enabled conversational journaling systems. The findings from our two empirical studies highlight key design and implementation challenges that could undermine the utility of conversational journaling, while also showcasing its potential to enhance data collection and improve patient experience. Additionally, we contribute to the field of healthcare by demonstrating the potential of conversational journaling to improve patient care and engagement.

\subsection{Facilitating the Perception of Social Presence}
Seven of our Study II participants (RP1-3, RP6, NP7-9) expressed that in their interactions with \toolname{}, the system came across as being sensitive, caring, and attentive.  According to the social presence theory~\cite{biocca2003toward}, attributing such qualities to a system by a person indicates the feeling or perception of interacting with another human, even with the knowledge that the other ``human'' is a computerized system. Prior research~\cite{hew2023using, jin2023social, tsai2021chatbots} suggest a positive relationship between the perception of social presence and prolonged engagement with conversational systems. We posit several aspects of \toolname{} collectively contributed to instilling a sense of social presence in our participants. The first aspect is the incorporation of the principles of cooperative conversation in our dialogue design. Cooperative conversation, as outlined by Gricean Maxims, emphasizes the importance of meaningful, relevant, truthful, and clear communication. Adhering to these principles fosters a more natural and engaging interaction that more closely mirrors human-human communication. This, in turn, enhances the participants' perception of the conversational system as a social entity rather than just a machine. The second aspect is the inclusion of empathetic responses. Empathy allows the system to recognize and respond to the participants' emotions, fostering a deeper emotional connection. Feeling understood and validated on an emotional level is then likely to enhance the participants' perception of the system as a social being. The third aspect is personalization. Acknowledging each individual's unique situation makes interactions between a participant and the system feel less generic and more personal. Prior research~\cite{hew2023using, blumel2023personal, laban2020effect} suggest that personalization notably enhances social presence in conversational systems by making interactions feel more relevant, engaging, and tailored to individual users.

We posit that the synergy created by the combination of cooperative conversations, empathetic responses, and personalization amplified the perception of social presence in participants' interactions with \toolname{}. Utilizing a similar design is likely to promote the perception of social presence in other conversational systems aimed at different patient populations and healthcare applications. Personalization can make each patient feel uniquely cared for, while empathy and cooperative dialogue can foster a sense of connection and trust.

\subsection{Enabling Accurate Intent Identification}
Accurate intent identification is at the core of enabling sustained conversational journaling. Several critical aspects of journaling, such as the generation of relevant follow-ups, proper empathetic responses, and reliable personalization, depend on accurately identifying intents and detecting the emotional tone of the patient's input. Outcomes of Study I suggest that utilizing traditional machine-learning models, such as decision trees and support vector machines~\cite{breiman1984classification,cortes:svm}, is likely to yield unsatisfactory intent identification outcomes. These models often lack the flexibility to handle the complexity and variability of natural language. Training specialized models tailored for healthcare could possibly improve performance. However, it would require large annotated datasets and substantial expertise in both machine learning and medical domains. The models might also face difficulties in scalability, maintaining context, and ensuring high accuracy and reliability. In Study II, we transitioned to utilizing an LLM, OpenAI's GPT-4, for intent identification, resulting in significant gains in intent identification accuracy (average of 99\%). The performance gain was not only for identifying intents in single-statement utterances but also in complex and convoluted bulk journal entries. Utilizing GPT-4 also notably improved the system's performance in properly personalizing interactions based on individual patient profiles and generating empathetic responses. These outcomes suggest that, from a practical standpoint, LLMs may currently be the best candidates for handling intent identification in conversational journaling systems~\cite{seo2024chacha}.

While promising, further research is required to learn best practices for integrating LLMs into conversational journaling systems for healthcare. 
Based on the results from Study II, personalization and flexibility are critical considerations for integrating LLMs into healthcare systems, especially for PwPD, whose symptoms and experiences can vary widely. Previous studies have also highlighted the need for personalized self-tracking devices and robust input mechanisms to improve clinician-patient communication and facilitate more effective treatment adjustments~\cite{mcnaney2020future, nunes2019agency, vafeiadou2021self}.
In addition, these models are resource-intensive, requiring substantial computational power to operate.

While previous research has emphasized the importance of privacy and data anonymity in healthcare systems~\cite{mcnaney2022exploring}, there are currently no established privacy and security recommendations and protocols that can guide developers in the privacy-aware use of LLMs. To safeguard participants' privacy, we ensured that no personal identifiers, such as the patient's name or age, were included in the conversation history shared with the API. While this measure addresses immediate privacy concerns, the regulatory landscape around using LLMs in healthcare is still evolving. 
\rr{In addition, free-form journaling in a home setting may unintentionally reveal sensitive patient information, including detailed personal experiences, relationships, or specific locations that could lead to identification. To address this challenge, systems like} \toolname{} \rr{could incorporate real-time feedback mechanisms that detect and flag potentially sensitive or unnecessary information}~\cite{ngong2024protecting} \rr{before passing patient inputs to the LLM. Such mechanisms would help patients rephrase or remove unnecessary personal details, promoting privacy-aware communication and aligning with evolving privacy standards in healthcare.} Further research is needed to thoroughly investigate the legal and ethical implications of deploying these technologies in medical settings.

\subsection{Supporting Ease of Access and Use}
The utility of our system was evident in the feedback from six of our participants (RP1, RP3, RP6, NP7-9), who emphasized that using voice as an input modality enhanced their experience by providing ease of access and use in comparison to writing or typing their journal entries.  Participants like RP1 and NP8 highlighted the convenience of speaking rather than typing, which made the journaling process more seamless and less tedious. Hence, using voice as the input modality can play a role in encouraging sustained journaling. Voice interaction can lower the accessibility barrier, especially for individuals who may have difficulty typing due to motor impairments~\cite{masina2020investigating}.  This ease of access can encourage more frequent entries, leading to richer data collection. Additionally, the use of voice interactions makes the system feel more engaging and conversational~\cite{reicherts2022s}, which can increase user satisfaction and adherence to the journaling process. 

While voice input offers several benefits for conversational systems, it is not universally accessible for everyone. Individuals with speech or hearing impediments or those who are hesitant or uncomfortable vocalizing their conditions may find voice journaling unsuitable~\cite{kane:2009}. Privacy is also another concern. Spoken words can be overheard by others. 
For example, participants in our study reported difficulties such as speaking softly due to Parkinson's (RP2), feeling self-conscious about discussing symptoms in front of others (RP3), and difficulties with voice recognition due to enunciation issues (NP8).
Voice data is innately more identifiable than written text, and the mechanisms by which it is stored and processed might be vulnerable to unauthorized access~\cite{kocaballi2022design}. 
Hence, when designing new technologies, we need to consider ways to incorporate a multimodal input approach that encompasses text input alongside voice, offering a more inclusive solution. This would address the needs of individuals with speech or hearing impairments and cater to those who prefer typing for privacy or other personal reasons. By offering flexibility and choice in input modalities, conversational journaling systems can enhance accessibility and cater to diverse user preferences. Future research can also investigate complementary technologies, such as sound masking~\cite{tung2019exploiting} in which a specially tuned ambient sound targets the same frequency as human speech, reducing its intelligibility.

\subsection{Fostering Mixed-Initiative Journaling}
Based on the feedback from several participants (RP2-4, RP6, NP8) who requested reminders to help them stay consistent with journaling, it is probable that for sustained engagement, we should shift from a purely reactive model, where journaling is initiated solely by the patient, to a mixed-initiative model in which both patient and system can initiate the conversation. A mixed-initiative approach can balance user autonomy with structured guidance to promote regular journaling and provide valuable health insights~\cite{bentley2013power}. For instance, the system could send reminders via email, text, or even voice notifications when a certain amount of time has passed without a journaling session. 
Additionally, these reminders can be individualized and personalized so that users can select their preferred notification modes. Personalized notifications for journaling have proven to be effective in healthcare settings. Chen et al.~\cite{chen2024investigating} for instance, investigated the use of push notifications in mobile applications to assist with weight management tracking and found that personalized notifications not only prompted timely responses but also facilitated the collection of more thorough health data.

Implementing such a proactive model not only helps in capturing critical health data more consistently but also empowers patients by actively involving them in their care process. Such proactive models align with the concept of persuasive technology in HCI~\cite{fogg2002persuasive}, in which systems are designed to positively influence user behaviors, usually through reminders and feedback. For instance, PDMove~\cite{zhang2019pdmove} uses smartphone sensors to monitor gait variability and send reminders to Parkinson's patients about missed or mistimed medication doses. By fostering a collaborative relationship between the user and the system, we can enhance the overall effectiveness of the journaling tool and contribute to better health outcomes.\enlargethispage{12pt}

\subsection{Enabling Analysis and Reflection}
To fully realize the potential of tools like \toolname{} and encourage sustained use, it is critical to complement data collection with robust data analysis. Providing patient and medical expert-facing analytics interfaces can enhance the utility of the collected data, supporting both PwPD and those involved in their personal and medical care. Participants (RP4 and NP7) in our study suggested the value of accessing and reviewing their journaling data, which could help them recognize patterns and make informed decisions about their health management. 

One practical application of deriving clinically relevant insights from the collected journal data is to monitor symptom fluctuations and medication effectiveness over time. Detailed records of motor symptoms and non-motor symptoms can help clinicians identify patterns related to medication timing and efficacy. For instance, if a patient consistently reports increased tremors before their next medication dose, this could indicate the need for adjusting dosing intervals or exploring alternative therapies. Neurologists in our study highlighted that such granular data could facilitate more precise and personalized treatment adjustments. These journaling systems can also be integrated with Electronic Health Records (EHRs) and secure data-sharing protocols~\cite{sun2011hcpp} to ensure journal data is readily accessible to healthcare providers before clinical consultations.

For patients, visualizing their journal data can enhance self-awareness and support self-management. Recognizing correlations between specific activities or environmental factors and symptom severity can allow patients to make lifestyle changes. For example, if a patient notices that stress exacerbates their symptoms, they may consider stress-reduction techniques. This process of making lifestyle changes based on self-awareness is consistent with the HCI concept of personal informatics, where self-tracking supports behavior change~\cite{li2010stage}.

By integrating advanced analytics, journaling tools can offer real-time feedback, trend analysis, and personalized recommendations, making the system engaging and valuable for users. This comprehensive approach can motivate patients to consistently log their data and ensure that the information collected is effectively utilized to improve health outcomes.

\subsection{Going Beyond Parkinson's Disease}

Many of \toolname{}'s design considerations and features could be adapted or adopted into the development of conversational journaling systems for other chronic conditions, such as diabetes or heart disease, where patients are often encouraged by healthcare providers to track their health data~\cite{ramesh2024data, rajabiyazdi2021communicating, rajabiyazdi2020exploring}. For system designers, our work provides key insights into the design principles that make conversational journaling systems effective. Prioritizing personalization and empathetic design can significantly enhance user engagement and data quality. Additionally, incorporating features that support in-context learning and real-time updates can make the system more flexible and responsive to user needs. These design considerations are crucial for developing AI systems that are not only technically proficient, but also user-friendly and impactful.


%% file: new_sections/8-limitations.tex
\section{Limitations}
Our results indicate that \toolname{} could be effective for journaling among PwPD. However, some limitations affect the scope of our findings and study operations. Both studies were conducted over two weeks, providing valuable initial insights into technical performance and participant experiences. However, this relatively short timeframe may not fully capture the long-term sustainability of engagement or the ongoing benefits of conversational journaling. 
Furthermore, while we attempted to support participant engagement by sending a reminder email every two days, this level of outreach may not have been sufficient to sustain long-term participation. Additionally, the passive nature of email reminders may not have effectively prompted action, especially for PwPD experiencing cognitive difficulties or fatigue. Future research should explore more robust engagement strategies, such as personalized reminders, adaptive messaging based on user behavior, or incorporating more interactive and multimodal reminders (e.g., text messages, app notifications, or phone calls) to better support the continued use of self-reporting systems in healthcare contexts.
Despite recruiting participants with diverse socio-demographic backgrounds, the sample sizes for Studies I (N=8) and II (N=9) were still limited. Additionally, the current symptoms included in \toolname{} may not fully encompass the broad range of experiences PwPD encounters. While the second iteration expanded symptom coverage, future refinements should include a wider range of symptoms to improve applicability. As a next step, we plan to deploy \toolname{} in a longitudinal study across multiple medical institutions to assess its scalability and generalizability. Furthermore, using voice as an input modality may be challenging for PwPD with speech and voice-related difficulties. Hence, we plan to implement a bi-modal input system supporting both text and voice to better accommodate user preferences and abilities.

%% file: new_sections/9-conclusion.tex
\section{Conclusion}
This work contributes to the growing body of research on AI-enabled conversational journaling systems, with \toolname{} representing a novel approach to exploring how AI can support semantically rich patient-generated health data collection. Our findings suggest that personalized, voice-enabled interactions, grounded in Gricean conversational principles of cooperation—such as relevance, clarity, and informativeness—can enhance patient engagement, particularly for PwPD, while addressing accessibility challenges related to motor impairments. Additionally, the system's ability to generate relevant follow-up questions plays an important role in eliciting detailed and contextually valuable information from patients, further enriching the data. Although our study focused on a specific population, the results point to the broader applicability of AI-driven conversational tools in chronic disease management. The integration of proactive engagement strategies, such as automated reminders, may support sustained journaling behavior, and the inclusion of caregiver and clinician interfaces could improve the clinical utility of the data collected. In summary, \toolname{} demonstrates the potential of AI in this domain, but further research is needed to explore its scalability, long-term efficacy, and adaptability to other chronic conditions. These insights can help inform the future development of AI-driven systems aimed at enhancing patient care and data quality in healthcare.